\documentclass[11pt,letterpaper]{article}

\usepackage{latexsym,multirow}
\usepackage{amssymb,amsmath, bm}
\usepackage{graphicx}
\usepackage[color,all,import,arrow]{xy}
\usepackage{enumerate}
\usepackage{caption}

\usepackage[T1]{fontenc}
\usepackage{lmodern}

\usepackage{natbib}
\usepackage[pdftex, bookmarksopen=true, bookmarksnumbered=true,
pdfstartview=FitH, breaklinks=true, urlbordercolor={0 1 0}, citebordercolor={0 0 1}]{hyperref}
\usepackage{colortbl}
\usepackage{subfigure}
\usepackage{float}

\usepackage{dcolumn}
\newcolumntype{.}{D{.}{.}{-1}}
\newcolumntype{d}[1]{D{.}{.}{#1}}
\usepackage{theorem}
\theoremstyle{plain}
\theoremheaderfont{\bfseries}
\newtheorem{assumption}{Assumption}

\newtheorem{definition}{Definition}

\newtheorem{theorem}{Theorem}

\newtheorem{lemma}{Lemma}

\newcommand{\ind}{\mbox{$\perp\!\!\!\perp$}}

\usepackage{rotating}

\usepackage[compact]{titlesec}

\usepackage{booktabs}
\usepackage{threeparttable}

\allowdisplaybreaks

\newcommand\spacingset[1]{\renewcommand{\baselinestretch}%
{#1}\small\normalsize}

\newcommand{\blind}{0}

\newcommand*{\QEDB}{\hfill\ensuremath{\square}}

\newcommand{\bone}{\mathbf{1}}

\newcommand{\pr}{\mathbb{P}}
\newcommand{\var}{\textnormal{var}}
\newcommand{\cov}{\textnormal{cov}}

\newcommand{\E}{\mathbb{E}}

\begin{document} 

\newcommand{\tit}{Connections as treatment: causal inference with edge interventions in networks
}
%
%
\spacingset{1.25}

\if0\blind

{\title{\bf\tit}

  \author{Shuli Chen\thanks{School of Mathematics,
      Sun Yat-sen University, Guangzhou,  Guangdong 510275, China. Email:  \href{mailto:chenshli27@mail2.sysu.edu.cn}{chenshli27@mail2.sysu.edu.cn}}
    \and
  Jie Hu\thanks{School of Mathematical Sciences,
      Xiamen University, Xiamen,  Fujian }
  \and
     Zhichao Jiang\thanks{School of Mathematics,
      Sun Yat-sen University, Guangzhou,  Guangdong 510275, China. Email:
      \href{mailto:jiangzhch7@mail.sysu.edu.cn}{jiangzhch7@mail.sysu.edu.cn} }
}

\date{
\today
}

\maketitle

}\fi

\if1\blind
\title{\bf \tit}

\maketitle
\fi

\pdfbookmark[1]{Title Page}{Title Page}

\thispagestyle{empty}
\setcounter{page}{0}

\begin{abstract}
Causal inference has traditionally focused on interventions at the unit level.  In many applications, however, the central question concerns the causal effects of connections between units, such as transportation links, social relationships, or collaborative ties. We develop a causal framework for edge interventions in networks, where treatments correspond to the presence or absence of edges. Our framework defines causal estimands under stochastic interventions on the network structure and introduces an inverse probability weighting estimator under an unconfoundedness assumption on edge assignment. We estimate edge probabilities using exponential random graph models, a widely used class of network models. We establish consistency and asymptotic normality of the proposed estimator.  Finally, we apply our methodology to China’s transportation network to estimate the causal impact of railroad connections on regional economic development.

\noindent {\bf Keywords:} design-based framework; interference; inverse probability weighting; random graph model; stochastic intervention
\end{abstract}


\clearpage
\spacingset{1.5}

\section{Introduction}
One of the key drivers in the advancement of causal inference has been the effort to accommodate increasingly diverse forms of treatment. Early methods mainly focused on binary treatments, as they provided a simple setting to develop and formalize the foundational ideas of causal inference \citep[e.g.,][]{rosenbaum1983central,imbens2015causal,ding2024first}. Subsequent developments have extended this framework to more complex treatment types, including categorical, continuous, functional, and point process data \citep[e.g.,][]{imbens2000role,imai2004causal,kennedy2017non,li2019propensity,zhang2021covariate,papadogeorgou2022causal,jiang2023instrumental,tan2025causal}. Despite this progress, most of the causal inference research continues to focus on treatments assigned at the unit level. However, in many practical settings, the primary object of intervention is not the unit itself, but the relationship between units. For example, \citet{sun2020displaying} examine how highlighting shared attributes influences friendship formation in online social networks, and \citet{gonzalez2016paving} investigate the effect of paving the streets on household consumption. 
These studies consider interventions on connections but do not provide a formal causal framework for defining or identifying edge-level effects.

We develop a design-based causal inference framework for edge interventions in undirected networks, where the treatment is defined by the presence or absence of edges between units. The treatments for all pairs of units collectively form a treatment matrix, which is the network’s adjacency matrix.
We define potential outcomes as functions of the network’s adjacency matrix, so that a unit’s outcome may depend not only on the edges directly incident to it but also on other edges in the network. This formulation generalizes the standard notation used in the interference literature, which typically allows a unit’s outcome to depend on the treatment assignments of other units \citep[e.g.,][]{manski1993identification,sobel2006randomized,hudgens2008toward,aronow2017estimating,savje2024causal}.
To make inference feasible, we introduce a local interference assumption that restricts the dependence of a unit’s potential outcome on the edges within a neighborhood. This extends prior local interference assumptions in network settings to the context of edge-level interventions  \citep{Toulis2013Estimation,sussman2017elements,karwa2018systematic,jagadeesan2020designs,awan2020almost,forastiere2021identification,forastiere2022estimating,leung2022causal,belloni2025neighborhood,gao2025causal,owusu2025randomization,wang2025design}.

Edge interventions present unique challenges for defining and identifying causal effects. First, the set of edges within each unit’s neighborhood may vary, making it difficult to define average potential outcomes under a single, fixed treatment condition. Second, the commonly invoked overlap assumption is often violated because of the large number of possible treatment values. To address these challenges, we define causal estimands under stochastic interventions, which enable specification of high-dimensional treatment assignment mechanisms through low-dimensional parameters  \citep[e.g.,][]{kennedy2019nonparametric,papadogeorgou2019causal,barkley2020causal, diaz2020causal,papadogeorgou2022causal,diaz2023nonparametric,wu2024assessing,zhou2024estimating,lee2025efficient}.

We establish identification of the proposed causal estimands under a generalized ignorability assumption tailored to edge interventions and develop corresponding inverse probability weighting (IPW) estimators. To implement the IPW estimators, we estimate edge probabilities by modifying the exponential random graph model (ERGM) with sample space constraints. These constraints prohibit edges between distant units, thereby reflecting the practical impossibility of direct connections in our application. 
We further provide conditions for the consistency and asymptotic Normality of the estimators. These conditions impose restrictions on the neighborhood size under the local interference assumption and on the dependence structure among edges in the network. Building on these results, we propose valid inference procedures.

We apply the proposed framework to evaluate how inter-city rail transportation connections affect local economic outcomes in China. In this application, the treatment is assigned to pairs of cities rather than individual cities. This naturally leads to a network representation, where each edge encodes a treatment status.
Much of the existing literature on the effect of transportation infrastructure simplifies this structure by defining treatment at the unit level, typically as a binary indicator for whether a city is connected to a major transportation network \citep[e.g.,][]{faber2014trade,banerjee2020road}. Such simplification obscures the relational nature of the treatment and limits the ability to capture how changes in connectivity affect both directly and indirectly connected cities.  Our analysis instead treats inter-city connections as edge-level treatments, demonstrating how the proposed methodology can be used to study changes in economic outcomes under stochastic increases or decreases in connection probabilities.

The remainder of the paper is organized as follows.  Section~\ref{sec::setup} presents the causal framework for edge interventions. Section~\ref{sec::method} develops the identification results, proposes the IPW estimators, describes the modified ERGM for estimating edge probabilities, and establishes asymptotic properties. Section~\ref{sec::sim} reports simulation studies, and Section~\ref{sec::empirical} applies the proposed method to analyze the data on China’s inter-city rail network. Section~\ref{sec::conclusion} concludes.

\section{A causal framework for edge interventions}
\label{sec::setup}
\subsection{Setup}
We consider a finite population setting that conditions on all potential outcomes and covariates, treating the treatment assignment as the only source of randomness.  This design-based framework is commonly employed in causal inference research \citep[e.g.,][]{imbens2015causal,aronow2017estimating,abadie2020sampling,leung2022causal,savje2024causal}. We focus on edge interventions in unweighted, undirected networks with $n$ units indexed by $\mathcal{N}=\{1,\ldots,n\}$. The treatment is represented by the network's adjacency matrix $\bm{A}=(A_{ij})^n_{i,j=1}$, where each entry $A_{ij}$ is a binary indicator of whether a connection exists between units $i$ and $j$. 
Self-connections are excluded so that $A_{ii}=0$ for all $i \in \mathcal{N}$.
The set of all such $n\times n$ binary adjacency matrices is denoted by $\mathcal{A}$. Each unit $i \in \mathcal{N}$ is associated with an observed outcome $Y_i$ and a covariate vector $X_i$. The covariate vector $X_i$ may include both unit-level characteristics and features related to its connections with other units. We denote the collection of all covariates by  $\bm{X}= \left( X_1,\ldots, X_n \right)$. 

We define the potential outcome for unit $i$ as $Y_i(\bm{a})$, representing the outcome that would be observed for unit $i$ if the network structure were set to $\bm{a}\in \mathcal{A}$.  This notation implies that $Y_i(\bm{a})$ depends not only on the edges directly connected to unit $i$, but also on the broader network structure, including edges between other units. As a result, each unit has $2^{n(n-1)/2}$ potential outcomes, which makes causal inference infeasible without additional assumptions. 

We impose a restriction on the dependence of potential outcomes on the network structure. Specifically, we assume that the potential outcome of unit $i$ depends only on the connections within a subnetwork formed by a subset of units. We refer to this subset of units as the exclusion neighborhood, denoted by $\mathcal{N}_i\subset \mathcal{N} $, which includes unit $i$ itself. This neighborhood is defined as the minimal set of units whose connections may influence $Y_i$. Thus, changes to the network outside $\mathcal{N}_i$ have no effect on $Y_i$. Let $\bm{A}_i=(A_{kl}:k,l\in \mathcal{N}_i)$
be the adjacency matrix of the subnetwork over $\mathcal{N}_i$, and let $\mathcal{A}_i$ be the set of all such binary adjacency matrices. We formalize this restriction in the following assumption.
\begin{assumption}(Local interference).
\label{asm::local}
For each $i\in \mathcal{N}$, $Y_i(\bm{a})=Y_i(\bm{a}^{\prime})$ if $\bm{a}_i=\bm{a}^{\prime}_i$ for any $\bm{a},\bm{a}^{\prime}\in \mathcal{A}$.
\end{assumption}
Assumption~\ref{asm::local}
implies that the outcome of unit $i$ depends solely on the network structure within its exclusion neighborhood $\mathcal{N}_i$.  Under this assumption, we can simplify $Y_i(\bm{a})$ as $Y_i(\bm{a}_{i})$. The exclusion neighborhood $\mathcal{N}_i$ is specified in advance and does not depend on the realized network. Its definition is based on external information or domain knowledge about which units may plausibly influence unit $i$. For example, in a transportation network, $\mathcal{N}_i$ may consist of cities within a certain geographical radius. In social networks, it may include individuals with similar demographic characteristics, interests, or organizational affiliation.

An alternative approach to restricting the interference structure is through exposure mapping, which assumes that the potential outcome depends on the treatments only through a specified function of the treatment values.  This approach facilitates the formal definition and identification of causal effects and are therefore widely used in the interference literature with unit-level treatments. 
Much of this work assumes that the exposure mapping is known in advance \citep[e.g.,][]{hudgens2008toward,manski2013identification,aronow2017estimating,basse2018analyzing,forastiere2022estimating,Bargagli2025HETEROGENEOUS}, while more recent research develops methods that allows for misspecified or unknown mappings \citep{savje2021average,leung2022causal,savje2024causal,viviano2024policy,gao2025causal}. 
For edge interventions, the exposure mapping idea can be generalized by assuming that the potential outcome depends on $\bm{a}_i$ only through certain summary statistics of this subnetwork. However, such an assumption imposes strong restrictions on how potential outcomes may depend on the subnetwork structure, and these restrictions generally cannot be verified from observed data. For this reason, we do not pursue this direction and instead rely solely on Assumption~\ref{asm::local}.

Without exposure mappings, a key complication arises because the size of the exclusion neighborhood $|\mathcal{N}_i|$ can vary across units. Consequently, the support of the local treatment $\bm{a}_i$ is unit-specific. This heterogeneity makes it difficult to define an average potential outcome under a single fixed treatment condition that applies to all units. Moreover, the large number of possible values of $\bm{a}_i$ often results in certain values having zero probability for some units, thereby violating the overlap assumption. In the following subsection, we will use stochastic interventions to define causal estimands.


\subsection{Causal estimands under stochastic interventions}\label{subsec::estimand}
Let $\delta$ denote a treatment strategy, defined as a probability distribution over the set of possible local treatment values for each unit $i$. Under this strategy, the treatment received by unit $i$ is a random variable $\bm{A}_i^\delta$ drawn from the distribution specified by $\delta$. The quantity  $\pr(\bm{A}_i^\delta = \bm{a}_i)$ represents the probability that unit $i$ receives a particular local treatment value $\bm{a}_i$ under $\delta$. We parameterize this probability through the following log odds ratio:
\begin{eqnarray}
\label{eq:intervention}
  \log \left\{ \frac{\pr(\bm{A}^{\delta}_i=\bm{a}_i)/\pr(\bm{A}^{\delta}_i=\bm{0}_i)}{\pr(\bm{A}_i=\bm{a}_i)/\pr(\bm{A}_i=\bm{0}_i)} \right\} &=& e(\bm{a}_i) \lambda_{\delta}
\end{eqnarray}
for any  $\bm{a}_i \in \mathcal{A}_i$,  
where $\bm{0}_i$ denotes the local treatment value with all entries equal to zero, 
$e(\bm{a}_i) = \sum_{k,j\in \mathcal{N}_i,k<j}a_{kj}$ is the number of edges in $\bm{a}_i$, and $\lambda_\delta$ is a pre-specified scalar parameter characterizing strategy $\delta$. Equation~\eqref{eq:intervention} specifies the odds ratio comparing the probability of $\bm{a}_i$ to that of $\bm{0}_i$ under the treatment strategy $\delta$ relative to the corresponding odds under the observed  distribution. 
Combined with the normalization constraint
 $\sum_{\bm{a}_i \in \mathcal{A}_i} \pr(\bm{A}^{\delta}_i=\bm{a}_i)=1$, this specification determines the entire probability distribution of $\bm{A}_i^\delta$.  

The parameter $\lambda_\delta$ acts as a global scaling factor for edge formation within the exclusion neighborhood of unit $i$. Consider two local treatment values $\bm{a}_i$ and $\bm{a}_i^\prime$ that differ only by a single symmetric entry, with $\bm{a}_i$ including an edge and  $\bm{a}_i^\prime$ excluding it. Then Equation~\eqref{eq:intervention} implies that 
\begin{eqnarray*}
  \log \left\{ \frac{\pr(\bm{A}^{\delta}_i=\bm{a}_i)/\pr(\bm{A}^{\delta}_i=\bm{a}_i^\prime)}{\pr(\bm{A}_i=\bm{a}_i)/\pr(\bm{A}_i=\bm{a}_i^\prime)} \right\} &=&\{e(\bm{a}_i)-e(\bm{a}_i^\prime)\}  \lambda_{\delta}\ = \ \lambda_{\delta}.
\end{eqnarray*}
Therefore, $\lambda_\delta$ measures the change in log odds  of adding a single edge under strategy $\delta$ relative to the observed distribution.
As a result, a positive $\lambda_\delta$ increases the probability of edge formation in the local treatment of unit $i$, while a negative $\lambda_\delta$ decreases it. 
When $\lambda_\delta=0$, the treatment strategy $\delta$ coincides with the observed edge distribution.

The specification in Equation~\eqref{eq:intervention} can be generalized to allow $\lambda_\delta$ to vary across units, which would enable differential emphasis on specific units such as economically or strategically important cities. However, this unit-specific parametrization requires specifying a separate parameter for each unit, which is often impractical and difficult to justify. For tractability, we focus on a scalar $\lambda_\delta$ shared across units.

 From Equation~\eqref{eq:intervention},  if $\pr(\bm{A}_i=\bm{a}_i)=0$ for some $\bm{a}_i$, then $\pr(\bm{A}_i^\delta=\bm{a}_i)=0$ as well.  In other words, the treatment strategy $\delta$ assigns zero probability to any local treatment value that is impossible under the observed distribution. This guarantees that the following overlap condition holds under our formulation of the treatment strategy.

\begin{assumption}(Overlap).
\label{asm::overlap}
There exists a constant $\alpha > 0$ such that
$ \pr(\bm{A}_i=\bm{a}_i)>\alpha \pr(\bm{A}_i^\delta = \bm{a}_i),$ for all $i\in \mathcal{N}$ and $\bm{a}_i \in \mathcal{A}_i$.
\end{assumption}
Assumption~\ref{asm::overlap} is weaker than the standard overlap assumption because it allows certain treatment values to have zero probability. It requires that the treatment strategy $\delta$ assign positive probability only to local treatment values with a non-zero probability in the observed distribution. This constraint limits the choice of $\delta$ but avoids the need to posit counterfactual treatment values that are impossible in the observed network. Similar overlap assumptions are commonly imposed in the stochastic intervention literature \citep[e.g.,][]{papadogeorgou2022causal,diaz2023nonparametric,wu2024assessing,zhou2024estimating}.

We now define the causal estimand of interest. For each unit $i$, the average potential outcome under treatment strategy $\delta$ is
\begin{eqnarray*}
    \theta_i^{\delta}&=& \E\left\{Y_i(\bm{A}_i^\delta)\right\}\ =\ \sum_{\bm{a}_i\in \mathcal{A}_i} Y_i(\bm{a}_i) \pr(\bm{A}_i^\delta = \bm{a}_i),
\end{eqnarray*}
which represents the expected outcome for unit $i$ when its local treatment is assigned according to strategy $\delta$. Averaging over all units yields the population-level average potential outcome
\begin{eqnarray*}
\theta^{\delta}&=&\frac{1}{n}\sum^n_{i=1}\theta^{\delta}_i.
\end{eqnarray*}
As a special case, we denote by $\theta^{0} $  the average potential outcome corresponding to the strategy with $\lambda_\delta=0$, under which $\bm{A}_i^\delta$ follows the same distribution as the observed treatment $\bm{A}_i$.
The quantity $\theta^{\delta}$ enables comparisons between treatment strategies through differences in average potential outcomes. 
For example, comparing $\theta^{\delta}$ with $\theta^{0}$ 
 assesses the  causal impact of shifting the distribution of local network connections from the observed distribution to that specified by strategy $\delta$.

\section{Identification and estimation}\label{sec:estimation}
\label{sec::method}
\subsection{Identification}
We extend the standard unconfoundedness assumption to the setting of edge interventions.
\begin{assumption}(Unconfoundedness).
\label{asm::unconfoundedness}
$\pr(\bm{A}\mid \bm{X}, \{Y_i(\bm{a}):\bm{a}\in\mathcal{A},i\in \mathcal{N}\})=\pr(\bm{A}\mid \bm{X}).$
\end{assumption}
Assumption~\ref{asm::unconfoundedness} states that, conditional on the covariates $\bm{X}$, the distribution of the adjacency matrix $\bm{A}$ does not depend on the potential outcomes. This requires that $\bm{X}$ include all covariates that jointly influence both the network structure and the outcomes. If relevant confounders are omitted from $\bm{X}$, then the assumption will be violated, leading to biased estimates. In such cases, sensitivity analyses are necessary to evaluate the robustness of the findings.

The following theorem establishes the nonparametric identification of $\theta^\delta$.
\begin{theorem}
\label{th::identification}
Under Assumptions~\ref{asm::local},~\ref{asm::overlap}, and~\ref{asm::unconfoundedness}, $\theta^\delta$ is identified by
\begin{eqnarray*}
\theta^{\delta}&=&\mathbb{E}\left\{\frac{1}{n}\sum_{i=1}^n\frac{\pr(\bm{A}^{\delta}_{i})}{\pr(\bm{A}_{i})}Y_i\right\}.
\end{eqnarray*}
\end{theorem}
Theorem~\ref{th::identification} shows that $\theta^\delta$ is equal to the expectation of a reweighted average of observed outcomes, where the weight $\pr(\bm{A}^{\delta}_{i})/\pr(\bm{A}_{i})$ represents the relative likelihood of observing local treatment $\bm{A}_i$ under strategy $\delta$ compared to the observed distribution. Assumption~\ref{asm::overlap} guarantees that the weights are well defined for all $\bm{a}_i$ with $\pr(\bm{A}_{i}=\bm{a}_i)>0$. 

 Theorem~\ref{th::identification} motivates the following unbiased estimator for $\theta^\delta$:
 \begin{eqnarray*}
 \label{eqn::unbiased}
     \frac{1}{n}\sum_{i=1}^n\frac{\pr(\bm{A}^{\delta}_{i})}{\pr(\bm{A}_{i})}Y_i.
 \end{eqnarray*}
However, this estimator is not directly computable because both
$\pr(\bm{A}_{i})$ and $\pr(\bm{A}^\delta_{i})$ are unknown. 
Therefore, we estimate these probabilities and then plug them into~\eqref{eqn::unbiased} to obtain the following Horvitz-Thompson inverse probability weighting (IPW) estimator:
\begin{eqnarray*}
   \hat{\theta}^{\delta}_{1}&=&\frac{1}{n}\sum_{i=1}^n\frac{\hat{\pr}(\bm{A}^{\delta}_{i})}{\hat{\pr}(\bm{A}_{i})}Y_i. 
\end{eqnarray*}
In finite samples, the estimated weights $\hat{\pr}(\bm{A}^{\delta}_{i})/\hat{\pr}(\bm{A}_{i})$ can exhibit high variability, especially when $\hat{\pr}(\bm{A}_{i})$ is close to zero. To mitigate this issue, we can use the following H\'ajek IPW estimator:
\begin{eqnarray*}\label{estimator:Hájek}  
\hat{\theta}^{\delta}_{2}&=&\left. \sum_{i=1}^n\frac{\hat{\pr}(\bm{A}^{\delta}_{i})}{\hat{\pr}(\bm{A}_{i})}Y_i\right/ \sum_{i=1}^n\frac{\hat{\pr}(\bm{A}^{\delta}_{i})}{\hat{\pr}(\bm{A}_{i})}.
\end{eqnarray*}
This estimator normalizes the weights in $\hat{\theta}^{\delta}_{1}$, thereby reducing the influence of extreme values and improving finite-sample stability.

Estimating these treatment probabilities is more challenging in the context of edge interventions than in conventional unit-level treatment settings. We describe our estimation approach in the next subsection.

\subsection{Estimation based on network modeling \label{subsec:modeling}}
 We estimate the treatment probabilities $\pr(\bm{A}_i)$ and $\pr(\bm{A}^\delta_i)$ using the exponential random graph model (ERGM), a widely used class of statistical models for network data \citep[e.g.][]{holland1981exponential,Frank1986}.  In its standard form, an ERGM specifies an exponential-family distribution for the adjacency matrix $\bm{A}$:
 \begin{eqnarray}
 \label{eq:ERGM model}
\pr_\eta(\bm{A}=\bm{a})&=&\frac1{\sum_{\bm{a}'\in \mathcal{A}}\exp\{\eta^\top g(\bm{a}',\bm{x})\}}\exp\{\eta^T g(\bm{a},\bm{x})\}
\end{eqnarray}
 for all $\bm{a}\in \mathcal{A}$, where $g(\bm{a},\bm{x})=(g_1(\bm{a},\bm{x}),\ldots,g_d(\bm{a},\bm{x}))^\top$ is a vector of network statistics,  and $\eta\in \mathbb{R}^d$ is the associated parameter vector. The network statistics $g(\bm{a},\bm{x})$ capture structural and covariate-dependent features of the network, such as the total number of edges, node degrees, and higher-order patterns like triangles or $k$-stars.
Given an adjacency matrix $\bm{a}$, let $\bm{a}^{ij+}$ and $\bm{a}^{ij-}$ denote the adjacency matrices by setting the $(i,j)$ entry to $1$ and $0$, respectively, while keeping all other entries unchanged.  The ERGM in~\eqref{eq:ERGM model} implies the following expression for the conditional log odds of an edge between units $i$ and $j$: 
 \begin{eqnarray}
 \label{eq::logodds}
     \log\left\{\frac{\pr \left(A_{ij}=1\mid A_{uv}=a_{uv} \forall~ (u,v)\neq (i,j)\right)}{1-\pr \left(A_{ij}=1\mid A_{uv}=a_{uv} \forall~ (u,v)\neq (i,j)\right)} \right\}&=&\eta^T\{g(\bm{a}^{ij+},\bm{x})-g(\bm{a}^{ij-},\bm{x})\}.
 \end{eqnarray}
This expression shows that the log odds of forming an edge depends on how the network statistics change when the edge is toggled. Therefore, the choices of network statistics directly determine the types of dependencies among edges in the network, while the parameter vector $\eta$ quantifies their direction and strength.

 A common feature of ERGMs is that they assign strictly positive probability to every possible binary adjacency matrix, implying that any pair of units has a nonzero probability of being connected.
 This property makes standard ERGMs unsuitable for modeling road networks in our application, as certain city pairs are too far apart to feasibly support a direct connection within a two-hour travel window. 
 
 We adapt the standard ERGM by restricting the sample space of $\bm{A}$ to adjacency matrices that allow edges only between units within a specified neighborhood. For each unit $i$, let  $\mathcal{U}_i$ denote the set of units eligible to form an edge with unit $i$. By construction, if unit $j \in \mathcal{U}_i$, then $i \in \mathcal{U}_j$. The collection $\{\mathcal{U}_i, i \in \mathcal{N}\}$ defines the restricted sample space $\mathcal{A}_{\mathcal{U}} = \{(a_{ij}): j \in \mathcal{U}_i \ \forall\ (i,j) \}$, which contains all binary adjacency matrices consistent with the requirement that edges may only be formed between $i$ and the units in $\mathcal{U}_i$ for all $i$.
The constrained ERGM is then defined as 
\begin{eqnarray}\label{ERGM model with constraints}
\pr_\eta(\bm{A=a})&=&\frac{ \bm{1}(\bm{a}\in \mathcal{A}_{\mathcal{U}})\exp\{\eta^T g(\bm{a},\bm{x})\}}{\sum_{\bm{a}'\in \mathcal{A}}\bm{1}(\bm{a}'\in \mathcal{A}_{\mathcal{U}})\exp\{\eta^T g(\bm{a}',\bm{x})\}},
\end{eqnarray}
where $\bm{1}(\cdot)$ is the indicator function. This formulation preserves the exponential-family structure for networks in  $\mathcal{A}_{\mathcal{U}}$ while assigning zero probability to all others.
The neighborhood $\mathcal{U}_i$ should be specified a priori based on substantive knowledge or external constraints on feasible connections. In practice, such restrictions might be determined by geographical distance in transportation networks or by institutional or technological limitations in social networks. We emphasize that $\mathcal{U}_i$ denotes a pre-specified feasibility neighborhood that restricts the support of the network, and is distinct from the exclusion neighborhood $\mathcal{N}_i$, which governs how the outcome for unit $i$ may depend on edges in the network.

We estimate the parameter $\eta$ using the Markov chain Monte Carlo maximum likelihood estimation method. This approach was first proposed by \citet{geyer1992constrained} for complex exponential family models for dependent data and  was later applied to the ERGM estimation \citep[e.g.,][]{snijders2002markov,hunter2008ergm}. 
For our constrained ERGM, we modify the proposal distribution in the Metropolis–Hastings step so that all proposed networks satisfy the constraints, ensuring that the chain remains entirely within the restricted space.


We then compute the estimators based on the constrained ERGM.
Let $\bm{a}_{-i}$ denote the sub-adjacency matrix of $\bm{a}$ obtained by removing all edges within the exclusion neighborhood $\mathcal{N}_i$, and let $\mathcal{A}_{-i}$ be the set of all possible values of $\bm{a}_{-i}$.
The full adjacency matrix can then be expressed as $\bm{a} = (\bm{a}_i, \bm{a}_{-i})$.
Under the constrained ERGM in~\eqref{ERGM model with constraints}, the observed distribution of the  local treatment is
\begin{eqnarray*}
\pr(\bm{A}_i=\bm{a}_{i})&=&
\sum_{\bm{a}_{-i}\in \mathcal{A}_{-i}}\pr\left(\bm{A}=(\bm{a}_i,\bm{a}_{-i})\right)
\ =\  \frac{\sum_{\bm{a}_{-i}\in \mathcal{A}_{-i}}\bm{1}((\bm{a}_i,\bm{a}_{-i})\in \mathcal{A}_{\mathcal{U}})  \exp\{\eta^T g(\bm{a}_i,\bm{a}_{-i},\bm{x})\}}{\sum_{\bm{a}'\in \mathcal{A}}\bm{1}(\bm{a}'\in \mathcal{A}_{\mathcal{U}})\exp\{\eta^T g(\bm{a}',\bm{x})\}},
\end{eqnarray*}
which marginalizes over  possible values of $\bm{a}_{-i}$ in $\mathcal{A}_{-i}$. 
Under the stochastic intervention defined in~\eqref{eq:intervention}, the probability of the local treatment under the treatment strategy $\delta$ is 
\begin{eqnarray*}
\pr(\bm{A}_i^\delta=\bm{a}_{i})&=&
\frac{\sum_{\bm{a}_{-i}\in \mathcal{A}_{-i}}\bm{1}((\bm{a}_i,\bm{a}_{-i})\in \mathcal{A}_{\mathcal{U}})  \exp\{\eta^T g(\bm{a}_i,\bm{a}_{-i},\bm{x})+ e(\bm{a}_i)\lambda_\delta \}}{\sum_{\bm{a}'\in \mathcal{A}}\bm{1}(\bm{a}'\in \mathcal{A}_{\mathcal{U}})\exp\{\eta^T g(\bm{a}',\bm{x})+e(\bm{a}'_i)\lambda_\delta \}}.
\end{eqnarray*}

By substituting the ERGM estimate $\hat{\eta}$ into the expressions of $\pr(\bm{A}_i=\bm{a}_{i})$ and $\pr(\bm{A}_i^\delta=\bm{a}_{i})$, we obtain 
\begin{eqnarray*}
\pr_{\hat{\eta}}(\bm{A}_i=\bm{a}_{i})&=&
 \frac{\sum_{\bm{a}_{-i}\in \mathcal{A}_{-i}}\bm{1}((\bm{a}_i,\bm{a}_{-i})\in \mathcal{A}_{\mathcal{U}})  \exp\{\hat{\eta}^T g(\bm{a}_i,\bm{a}_{-i},\bm{x})\}}{\sum_{\bm{a}'\in \mathcal{A}}\bm{1}(\bm{a}'\in \mathcal{A}_{\mathcal{U}})\exp\{\hat{\eta}^T g(\bm{a}',\bm{x})\}},\\
 \pr_{\hat{\eta}}(\bm{A}_i^\delta=\bm{a}_{i})&=&
\frac{\sum_{\bm{a}_{-i}\in \mathcal{A}_{-i}}\bm{1}((\bm{a}_i,\bm{a}_{-i})\in \mathcal{A}_{\mathcal{U}})  \exp\{\hat{\eta}^T g(\bm{a}_i,\bm{a}_{-i},\bm{x})+ e(\bm{a}_i)\lambda_\delta \}}{\sum_{\bm{a}'\in \mathcal{A}}\bm{1}(\bm{a}'\in \mathcal{A}_{\mathcal{U}})\exp\{\hat{\eta}^T g(\bm{a}',\bm{x})+e(\bm{a}'_i)\lambda_\delta \}},
\end{eqnarray*}
We use the subscript $\hat{\eta}$ to emphasize that these probabilities are computed under the estimated ERGM parameter.
These probabilities yield the following IPW estimators:
\begin{eqnarray}
 \label{eqn::est1}   \hat{\theta}^\delta_1&=& \frac{1}{n}\sum_{i=1}^n \frac{\exp\{e(\bm{A}_i)\lambda_\delta\}}{\sum_{\bm{a}'_i\in \mathcal{A}_i}\exp\{e(\bm{a}'_i)\lambda_\delta\}\pr_{\hat{\eta}}(\bm{A}_i=\bm{a}'_{i})}Y_i,\\
  \label{eqn::est2}    \hat{\theta}^\delta_2&=& \left.\sum_{i=1}^n \frac{\exp\{e(\bm{A}_i)\lambda_\delta\}Y_i}{\sum_{\bm{a}'_i\in \mathcal{A}_i}\exp\{e(\bm{a}'_i)\lambda_\delta\}\pr_{\hat{\eta}}(\bm{A}_i=\bm{a}'_{i})}\right /  \sum_{i=1}^n \frac{\exp\{e(\bm{A}_i)\lambda_\delta\}}{\sum_{\bm{a}'_i\in \mathcal{A}_i}\exp\{e(\bm{a}'_i)\lambda_\delta\}\pr_{\hat{\eta}}(\bm{A}_i=\bm{a}'_{i})}.
\end{eqnarray}
If we treat the estimated ERGM parameter $\hat{\eta}$ as fixed, then the denominator in the weights of~\eqref{eqn::est1} and \eqref{eqn::est2} can be written as
\begin{eqnarray}
\label{eqn::denominator}   \sum_{\bm{a}'_i}\exp\{e(\bm{a}'_i)\lambda_\delta\}\pr_{\hat{\eta}}(\bm{A}_i=\bm{a}'_{i}) &=& \E_{\hat{\eta}}\left\{ \exp\{e(\bm{A}_i)\lambda_\delta\}\right\},
\end{eqnarray}
where $\E_{\hat{\eta}}$ denotes the expectation taken with respect to the estimated distribution $\pr_{\hat{\eta}}(\bm{A}_i)$.
 In practice, we first estimate $\hat{\eta}$ in the constrained ERGM and then use Monte Carlo simulation from the fitted model to approximate the expectation in \eqref{eqn::denominator}.

\subsection{Asymptotic properties}
We now examine the asymptotic properties of the proposed estimators. Since our IPW estimators depend on the estimated ERGM parameter $\hat{\eta}$, their asymptotic behavior relies on the properties of the ERGM estimator.
However, establishing asymptotic theory for ERGMs is notably challenging because of the complex dependence structure among edges. While some results have been developed under the assumption of independent edges,  the general asymptotic behavior of ERGM estimators remains poorly understood. For further discussion of these challenges, see \citet{schweinberger2020concentration} and \citet{schweinberger2020exponential}.

To make the asymptotic analysis tractable, we treat the estimated parameter $\hat{\eta}$ as fixed when studying the asymptotic behavior of our estimators. This effectively ignores the estimation uncertainty in the ERGM and is equivalent to viewing the data as arising from a conditional randomized experiment in which the treatment is generated from the constrained ERGM with parameter $\hat{\eta}$. Similar simplifications are common in the matching and covariate balancing literature and in studies where quantifying the uncertainty in estimated treatment probabilities is difficult \citep[e.g.,][]{hainmueller2012entropy,shook2022power,papadogeorgou2025causal}. We will use simulation studies to assess the impact of this simplification.

We focus on $\hat{\theta}^\delta_1$ for the asymptotic analysis; the results for $\hat{\theta}^\delta_2$ are analogous. From~\eqref{eqn::est1}~and~\eqref{eqn::denominator}, we can write
 \begin{eqnarray}
 \label{eqn::theta1-sum}  \hat{\theta}^\delta_1&=&  \frac{1}{n}\sum_{i=1}^n \frac{\exp\{e(\bm{A}_i)\lambda_\delta\}}{\E_{\hat{\eta}}\left\{ \exp\{e(\bm{A}_i)\lambda_\delta\}\right\}}Y_i(\bm{A}_i).
 \end{eqnarray}
The summands in~\eqref{eqn::theta1-sum} are not independent because $\bm{A}_i$ and $\bm{A}_j$ may be correlated for $i \neq j$.  If the exclusion neighborhoods $\mathcal{N}_i$ and $\mathcal{N}_j$ intersect, then $\bm{A}_i$ and $\bm{A}_j$ share common edges, inducing correlations between them. This source of dependence is well recognized in the network causal inference literature and is typically addressed by imposing restrictions on the size of the exclusion neighborhood. However,  such restrictions are generally not sufficient for edge interventions, where the edges themselves may be correlated. This contrasts with prior studies that assume independent unit-level treatments across units \citep[e.g.,][]{leung2022causal,gao2025causal}.  As a result, the dependence structure among the $\bm{A}_i$’s is more complex in our setting.
To account for this additional complexity, we impose the following assumption on the dependence among edges.


\begin{assumption}
\label{asm::dependence}(Local dependence)
For each unit $i\in \mathcal{N}$,  there exists a subset $\mathcal{R}_i\subset \mathcal{N}$, such that $\bm{A}_{i}\ind \bm{A}_{j}$, for all $j\notin\mathcal{R}_i $.
 \end{assumption}
Assumption~\ref{asm::dependence} states that the local treatment $\bm A_i$ is independent of other local treatments $\bm{A}_j$ outside the set $\mathcal{R}_i$. It is important to distinguish between the sets $\mathcal{N}_i$ and $\mathcal{R}_i$. The set $\mathcal{N}_i$ determines which edges directly influence the potential outcomes of unit $i$, whereas $\mathcal{R}_i$ characterizes the correlation structure of the network by specifying the set of units whose local treatments may influence the distribution of $\bm A_i$.

The structure of $\mathcal{R}_i$ depends on both $\mathcal{N}_i$ and the dependence structure of the edges induced by the ERGM. 
Under the Bernoulli random graph model where edges are mutually independent, $\bm{A}_i$ and $\bm{A}_j$ are independent whenever $\mathcal{N}_i$ and $\mathcal{N}_j$ share fewer than two units. In this case, we can define $\mathcal{R}_i = \{j: |\mathcal{N}_i\cap \mathcal{N}_j|\geq 2\}$. However, under more general ERGMs where global dependence among edges is present \citep{Frank1986, snijders2006new}, such a set $\mathcal{R}_i$ may not exist. To ensure the existence of a valid local dependence structure, one must restrict edge dependence to pre-specified subgraphs, as in the framework proposed in \cite{schweinberger2015local}.  
In our simulation and empirical analyses, we focus on the Bernoulli random graph model with sample space constraints. We present additional results under more complicated ERGMs with local dependence in the appendix.

The following theorem establishes the consistency of  $\hat{\theta}^\delta_1$ under a condition on $\mathcal{R}_{i}$.
\begin{theorem}
\label{Th::Consistency of theta}
Suppose Assumptions~\ref{asm::local}~to~\ref{asm::dependence} hold and the outcome is bounded. If $\sum_{i=1}^n |\mathcal{R}_i|=o_\pr(n^2)$, then $\hat{\theta}^\delta_1$ converges to $\theta^\delta$ in probability.
\end{theorem}
The assumption of bounded outcomes can be relaxed to a weaker condition requiring only that the covariance between outcomes $Y_i$ and $Y_j$ be uniformly bounded. We adopt the stronger condition here for simplicity, and it holds in our application.
The condition $\sum_{i=1}^n |\mathcal{R}_i|=o_\pr(n^2)$ constrains the overall strength of dependence across the population. It requires that dependence among units be sufficiently sparse: although each unit’s local treatment $\bm{A}_i$ may depend on the treatments of other units in $\mathcal{R}_i$, the total number of such dependencies must grow slower than $n^2$. If each $\mathcal{R}_i$ includes a constant fraction of the population, resulting in dense dependence among edges, then the condition is violated with $\sum_{i=1}^n |\mathcal{R}_i| = O_\pr(n^2)$.
The stated condition rules out such dependence structure and instead requires that most units interact with only a vanishingly small fraction of the population as $n$ increases.



The following theorem establishes the asymptotic normality of $\hat{\theta}_1^\delta$.
 \begin{theorem}
\label{Th::CLT}
Suppose that Assumptions~\ref{asm::local}~to~\ref{asm::dependence} hold, the outcomes are bounded, and the condition in Theorem \ref{Th::Consistency of theta} holds. 
If 
\[
\max\left[n^{-2} \left\{\var(\hat{\theta}^\delta_1)\right\}^{-3/2}\max_{i\in\mathcal{N}}|\mathcal{R}_i|^2, n^{-3/2}\left\{\var(\hat{\theta}^\delta_1)\right\}^{-1}\max_{i\in\mathcal{N}}|\mathcal{R}_i|^{3/2} \right]\to 0,
\]
then 
\begin{eqnarray*}
\frac{\hat{\theta}^\delta_1-\theta^\delta}{\sqrt{\var(\hat{\theta}^\delta_1)}}\xrightarrow{d} \mathcal{N}\left(0,1\right).
\end{eqnarray*}
\end{theorem}
Theorem~\ref{Th::CLT} imposes a rate condition on $\max_{i\in\mathcal{N}}|\mathcal{R}_i|$, which represents the maximum number of other units whose local treatments are correlated with that of a given unit. 
This condition requires a more stringent sparsity in the network dependence structure than the one required for consistency in Theorem~\ref{Th::Consistency of theta}. Specifically, while Theorem~\ref{Th::Consistency of theta} permits a small number of units to have large dependence neighborhoods, Theorem~\ref{Th::CLT} restricts maximum level of dependence for all units. This ensures that no single unit exerts disproportionate influence on the estimator’s variability, which is essential for establishing asymptotic normality.
Conditions of this form are common in the literature on network-based causal inference, where asymptotic Normality is established for estimators that aggregate dependent observations \citep{kojevnikov2021limit,leung2022causal,gao2025causal}.
We then consider variance estimation. Because the summands in~\eqref{eqn::theta1-sum} are correlated, the standard bootstrap is invalid. Instead, we use the modified dependent wild bootstrap  proposed by \citet{kojevnikov2021bootstrap}, which constructs bootstrap perturbations with a correlation structure that reflects the dependence in the original network data.

We generate the bootstrap sample as 
\begin{eqnarray*}
    Y^\ast_i &=& \hat{\theta}^\delta_1+\left[\frac{\exp\{e(\bm{A}_i)\lambda_\delta\}}{\E_{\hat{\eta}}\left\{ \exp\{e(\bm{A}_i)\lambda_\delta\}\right\}}Y_i(\bm{A}_i)-\hat{\theta}^\delta_1\right]W_{i}
\end{eqnarray*}
for all $i$, where $(W_1,\ldots, W_n)$ follows a multivariate Normal distribution with mean zero and covariance matrix $\Omega$.
The matrix $\Omega$ encodes the dependence across local treatments: 
\begin{eqnarray*}
 \Omega &= &(\omega_{ij}),\quad   \omega_{ij}\ =\ \frac{|\mathcal{R}_i\cap \mathcal{R}_j|}{n^{-1}\sum_{i=1}^n |\mathcal{R}_i|}.
\end{eqnarray*}
For each bootstrap sample, we compute $\theta^*_1 =n^{-1}\sum_{i=1}^n Y^\ast_i$,  and estimate the variance using the sample variance across bootstrap replications. \citet{kojevnikov2021bootstrap} shows that this bootstrap variance estimator can be computed directly from the original sample as 
\begin{eqnarray}
\label{eqn::varest} \frac{1}{n^2}\sum_{i,j\in \mathcal{N}} \omega_{ij}\left[\frac{\exp\{e(\bm{A}_i)\lambda_\delta\}}{\E_{\hat{\eta}}\left\{ \exp\{e(\bm{A}_i)\lambda_\delta\}\right\}}Y_i(\bm{A}_i)-\hat{\theta}^{\delta}_1\right]\left[\frac{\exp\{e(\bm{A}_j)\lambda_\delta\}}{\E_{\hat{\eta}}\left\{ \exp\{e(\bm{A}_j)\lambda_\delta\}\right\}}Y_j(\bm{A}_j)-\hat{\theta}^{\delta}_1\right].
\end{eqnarray}
Therefore, we apply~\eqref{eqn::varest} directly for variance estimation. Consistency of this estimator for the asymptotic variance follows from the results in \citet{kojevnikov2021bootstrap}.

\section{Simulation}
\label{sec::sim}
We conduct simulation to evaluate the finite-sample performance of the proposed method.
We assess the accuracy of the point and variance estimators, the validity of the bootstrap variance estimators, and the coverage rates of the confidence intervals based on the asymptotic Normality of the estimators.

We first generate two independent covariates $X_{i1}$ and $X_{i2}$ from the standard Normal distribution. Following \citet{leung2022causal} and \citet{gao2025causal}, we assign each unit a location in a two-dimensional Euclidean space, with both coordinates independently drawn from the uniform distribution on $[0,1]$. We compute the Euclidean distance $d_{ij}$ between units $i$ and $j$ and define the exclusion neighborhood $\mathcal{N}_i$ using the \(l\)-nearest neighbors rule: $\mathcal{N}_i$ consists of unit $i$ together with its $l-1$ nearest neighbors. The constrained sample space $\mathcal{U}_i$ in~\eqref{ERGM model with constraints} is defined as
$\mathcal{U}_i = \{ j : d_{ij} < 0.2 \}$. 

We generate the potential outcomes from the following linear model:
\begin{eqnarray*}
    Y_i(\bm{a}_i) &=& 1 + 2X_{i1} + 1.5X_{i2} + \sum_{j \in \mathcal{N}_i} a_{ij} (X_{j1} + X_{j2}) + \epsilon_i,
\end{eqnarray*}
where $\epsilon_i \sim N(0,1)$. This setup ensures that each unit’s potential outcome depends only on the edges within $\mathcal{N}_i$, thus satisfying Assumption~\ref{asm::local}. 

We fix the potential outcomes in the simulation and generate the edge interventions $\bm{A}$ from the constrained ERGM in~\eqref{ERGM model with constraints} with three network statistics
\begin{eqnarray*}
g_1(\bm{a}, \bm{x}) = e(\bm{a}), \quad 
g_2(\bm{a}, \bm{x}) = \sum_{1 \leq i < j \leq n} a_{ij} (x_{i1} + x_{j1}), \quad 
g_3(\bm{a}, \bm{x}) = \sum_{1 \leq i < j \leq n} a_{ij} (x_{i2} + x_{j2}),
\end{eqnarray*}
and set the associated parameter vector to $\eta = (-1.5,\, 0.5,\, -0.5)^\top$. The statistic $g_1$ corresponds to the total number of edges, and $g_2$ and $g_3$ incorporate two covariates. 
Under this specification, edges are independent conditional on covariates. In the appendix, we further explore settings with more complex ERGMs that allow for dependent edges.
Finally, we compute the observed outcomes from the generated edge interventions and the potential outcomes for each unit.

We consider three sample sizes $n = 300$, $1000$, $5000$, and vary the exclusion neighborhood size with $l = 3$, $5$, and $7$. To capture different levels of intervention intensity, we explore five values of the stochastic intervention parameter: $\lambda_\delta = -\log (2), -\log(1.5), 0, \log(1.5)$, and $\log(2)$. These correspond to odds ratios for adding one edge of 0.5, 0.67, 1, 1.5, and 2, respectively.
We report the results on $\hat{\theta}_2^\delta$ and present the results on  $\hat{\theta}_1^\delta$ in the appendix.

\begin{figure}[htp]
    \centering
    \setlength{\abovecaptionskip}{-0.1cm}
    \includegraphics[width=16cm]{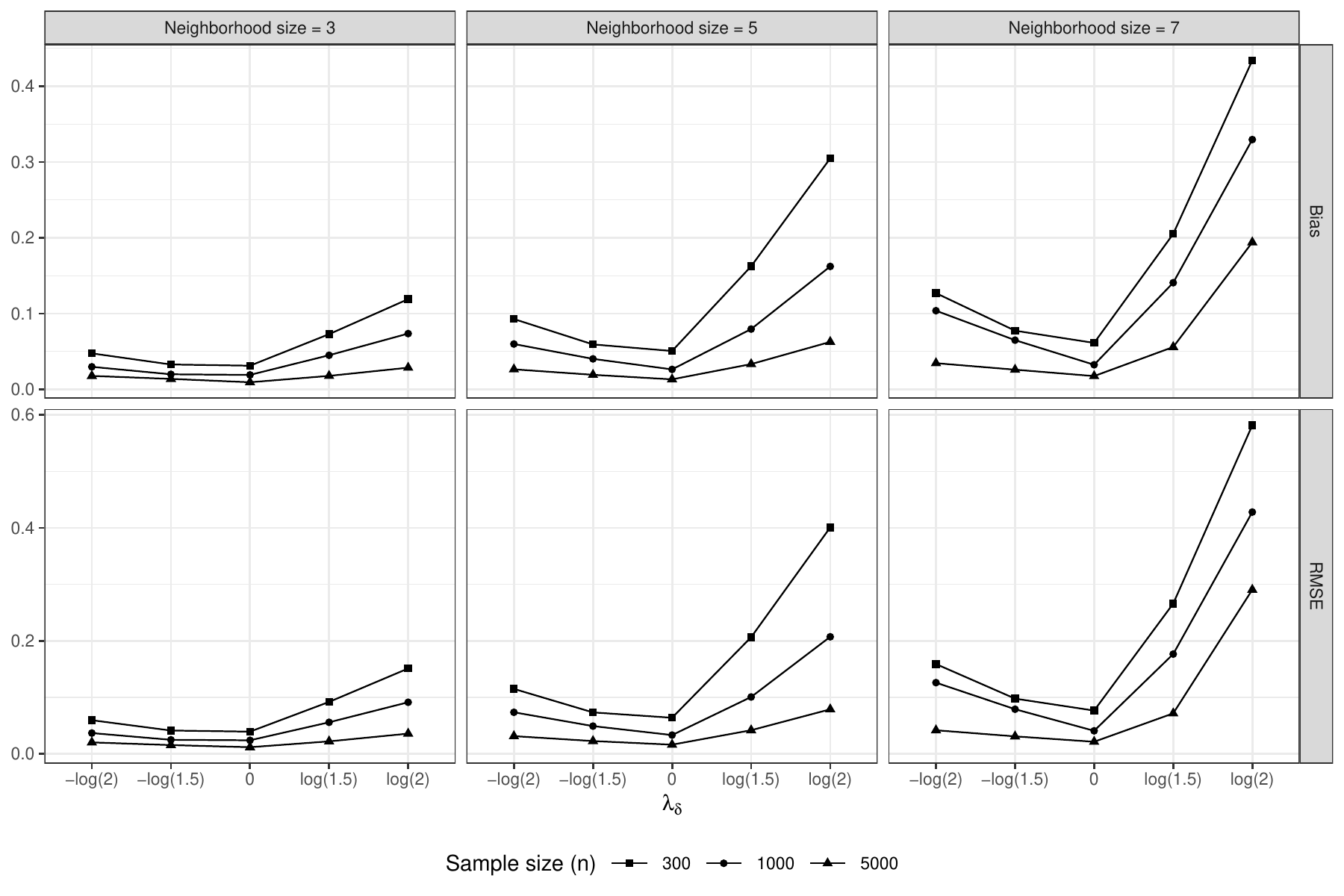}
\caption{Bias and RMSE of the Hájek estimator $\hat{\theta}^{\delta}_{2}$. The first row displays bias, and the second row displays RMSE. Each column represents a different exclusion neighborhood size $l \in \{3, 5, 7\}$.}
    \label{fig:sim_bias_rmse}
\end{figure}

We begin by evaluating the point estimation performance. Figure~\ref{fig:sim_bias_rmse} displays the biases and RMSEs of the Hájek estimator $\hat{\theta}^{\delta}_{2}$. Both metrics increase as $\lambda_\delta$ moves away from zero in either direction, reflecting greater variability under more extreme stochastic interventions. Larger exclusion neighborhoods lead to higher bias and RMSE, due to increased dependence among local treatments.
As expected, both bias and RMSE decline with increasing sample size, indicating improved estimator precision and supporting the asymptotic properties of the proposed method.

We next evaluate the performance of the variance estimator defined in~\eqref{eqn::varest}. We compare the estimated variances with the true sampling variances of the Hájek estimator $\hat{\theta}^{\delta}_{2}$, approximated using $1,000$ Monte Carlo replications.
We consider two versions of the true variance. The first incorporates the uncertainty from estimating the network model. In each replication, we generate edge interventions from the true ERGM and re-estimate the model parameters before computing the estimator. The resulting sampling variance thus reflects both outcome variation and model estimation uncertainty.
The second version conditions on the true network model parameters. In each replication, we still generate edge interventions from the true ERGM and compute the estimator using the known true parameter values, without re-estimating the model. This variance therefore treats the ERGM parameters as fixed and excludes uncertainty arising from model estimation.

\begin{figure}[htp]
    \centering
    \setlength{\abovecaptionskip}{-0.3cm}
    \includegraphics[width=17.5cm]{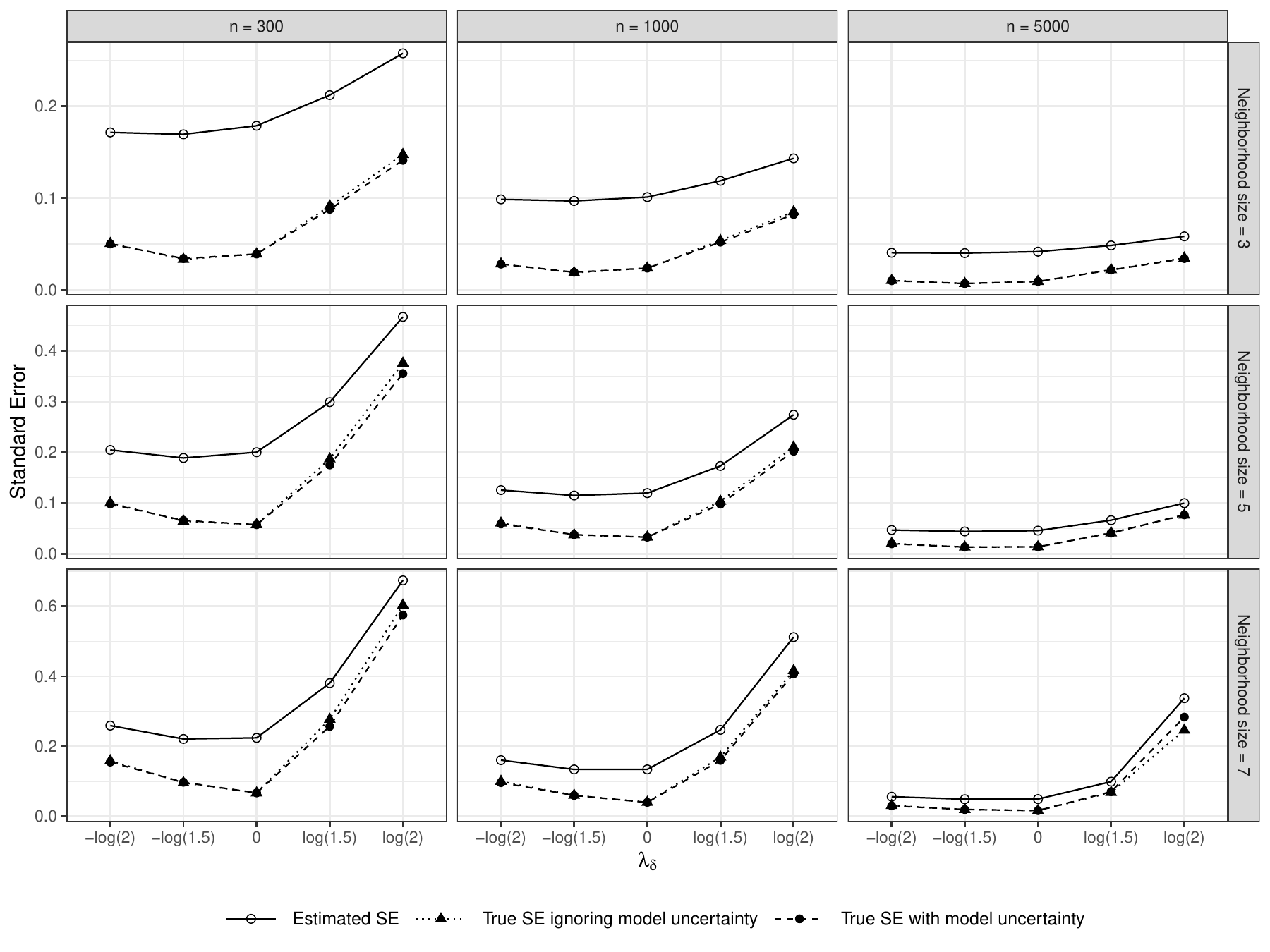}
    \caption{Comparison of estimated and true standard errors (SEs) for $\hat{\theta}^{\delta}_{2}$.
    Each panel shows standard errors with different values of $\lambda_\delta$.  Columns correspond to sample sizes ($n = 300, 1000, 5000$), and rows correspond to exclusion neighborhood sizes ($l = 3, 5, 7$). Solid lines indicate the estimated standard errors.
    Dashed lines represent the true standard errors accounting for ERGM estimation uncertainty.
  Dotted lines represent the true standard errors assuming the fitted ERGM is the true model. }
    \label{fig:sim_se}
\end{figure}

Figure~\ref{fig:sim_se} presents the results. Across all settings, the estimated standard errors exceed both versions of the true standard errors, indicating a conservative variance estimation. This conservativeness diminishes with increasing sample size and neighborhood size. One possible explanation is that the correlation structure used in variance estimation may overstate the actual dependence among edges. 
Moreover, the two versions of the true standard errors are generally close to each other, with the version that ignores model uncertainty being slightly larger. This empirical pattern supports the validity of our variance estimator. A more detailed theoretical understanding of this phenomenon would require refined probabilistic tools beyond the scope of this paper.

We conduct additional simulations in the appendix.  First, we examine the performance of $\hat{\theta}^\delta_1$ using the same data-generating process as in the main text. Second, we consider more complex network models by generating edge interventions from ERGMs with dependent edges. To ensure  that Assumption~\ref{asm::dependence} is satisfied, we incorporate the local dependence structure proposed by \citet{schweinberger2015local} into the ERGM specification.

\section{Empirical analysis}
\label{sec::empirical}
\subsection{Data and setup}
We now apply the proposed framework to evaluate the economic impact of inter-city rail connections in China. The primary dataset, compiled by \citet{ma2024distributional}, reports rail travel times in hours between 279 Chinese prefecture-level cities from 1994 to 2017. To ensure analytical consistency, we exclude four cities from the sample. Laiwu is dropped due to administrative restructuring, while Lhasa, Urumqi, and Karamay are excluded because their extreme geographical isolation renders them incomparable to the rest of the network. The final dataset thus consists of 275 prefecture-level cities. For each city pair, the dataset provides the shortest feasible rail travel time, calculated using a fast marching algorithm on a raster grid. Each pixel in the grid is assigned a design speed determined by construction vintage, infrastructure class, and terrain characteristics. The algorithm then simulates optimal travel paths across the grid to obtain the minimum travel time between city pairs.

\begin{figure}[t!]
    \centering
    \setlength{\abovecaptionskip}{-0.3cm}
    \includegraphics[width=\textwidth]{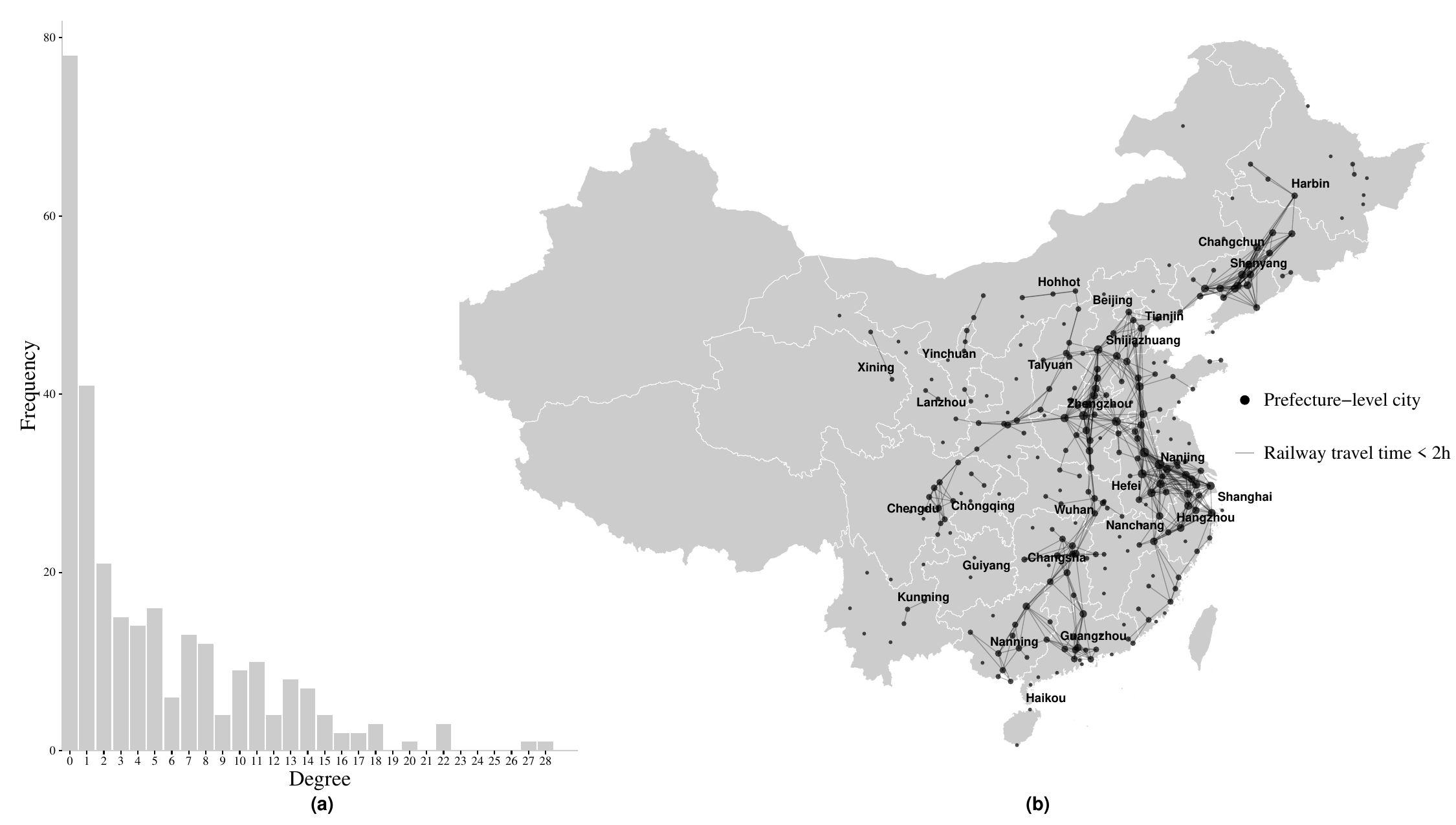}
    \caption{Panel (a) shows the degree distribution of the Chinese railway network. Panel (b) overlays the network on a geographic base map, where nodes represent prefecture-level cities and edges indicate rail connections with travel times under two hours. Node size is proportional to the number of direct connections. }
    \label{fig:rail_network}
\end{figure}

We make several simplifications to facilitate our analysis. First, we dichotomize travel times between city pairs using a two-hour threshold. Two cities are considered connected if the shortest rail travel time between them is less than two hours. This choice reflects both transport efficiency and commuting behavior. Prior research shows that rail becomes the dominant mode of travel when times fall below two hours \citep{jorritsma2009substitution,kroes2019substitution}, and commuting studies indicate that trips of this length remain acceptable for regular travel \citep{he2016tolerance,de2025makes}. 
Second, we focus on the 2017 rail network, the most recent year in the data.
These simplifications yield an undirected, unweighted network with 275 nodes and 647 edges. Figure~\ref{fig:rail_network}(a) shows the degree distribution of this network, which is highly skewed. Most cities have only a few direct connections, while a small number serve as major transportation hubs. This pattern reflects the selective and uneven expansion of China’s rail infrastructure, which has prioritized economically and strategically important regions. Figure~\ref{fig:rail_network}(b) maps the network onto a geographic base layer, where nodes represent cities and edges indicate rail connections with travel times under two hours.

Our goal is to assess how the rail connections in 2017 influence the tertiary industry value added in 2018, which we use as a proxy for the level of economic development. 
Prior studies have examined the effects of high-speed rail development on various aspects of regional growth in China \citep[e.g.,][]{wang2020impact,wang2022china}. These studies typically define treatment at the city level, using a binary indicator for whether a city is connected to the rail network, and often rely on structural outcome models. This unit-level framing overlooks the inherently relational nature of connectivity, and the reliance on outcome modeling introduces risks of misspecification.

\subsection{Estimation and results}

We address these limitations by developing a causal framework that explicitly accounts for edge interventions. In the sections that follow, we define causal estimands for edge interventions, establish identification conditions, and propose estimation procedures that handle the complexities of edge-level treatments.

We examine how average potential outcomes change under different treatment strategies, focusing on the contrast $\theta^\delta - \theta^0$ between a given stochastic intervention and the observed network. We consider four values for the intervention parameter: $\lambda_\delta = -\log(2), -\log(1.5), \log(1.5), \log(2)$. Negative values represent strategies that reduce the likelihood of local railway connections, while positive values reflect strategies that increase them.

We fit the constrained ERGM in~\eqref{ERGM model with constraints} to estimate the treatment probabilities. The network statistics include both network-only and covariate-dependent terms. The network-only term captures the total number of edges. For each covariate $X_{ip}$, we include two covariate-dependent terms: one that captures aggregate levels across connected cities, $\sum_{1\leq i<j\leq n}a_{ij}(x_{ip}+x_{jp})$, and another that measures similarity across edges. The similarity term takes the form $\sum_{1\leq i<j\leq n}a_{ij}|x_{ip}-x_{jp}|$ for continuous covariates, and $\sum_{1\leq i<j\leq n}a_{ij}\bm{1}(x_{i}=x_{j})$ for categorical covariates.
We collect a rich set of covariates from the 2016 edition of the China City Statistical Yearbook. These covariates include per capita GDP, local fiscal revenue, annual average population, highway freight volumes, civil aviation access, and administrative level. We also incorporate city-level terrain relief data from the Relief Degree of Land Surface Dataset of China \citep{you2018relief} to capture geographic constraints on infrastructure expansion.

We impose geographic constraints on the ERGM by limiting the set of possible connections based on physical distance. With slight abuse of notation, we define $\mathcal{U}_i = \{j: d_{ij} < d_u\}$, where $d_{ij}$ is the physical distance between cities $i$ and $j$, and $d_u$ is a distance cutoff. This constraint reflects practical transportation limits: city pairs located too far apart to be reached within two hours by rail are unlikely to form a direct connection. Given that China’s high-speed rail typically operates at a maximum speed of 400 km/h, we set $d_u = 800$ km as the baseline cutoff. To evaluate the robustness of our results, we conduct a sensitivity analysis in the appendix by varying the assumed travel time. Specifically, we consider cutoffs corresponding to 1.5-hour and 2.5-hour travel times, which translate to $d_u = 600$ km and $d_u = 1000$ km, respectively.

We define the exclusion neighborhood $\mathcal{N}_i$ of city $i$ as the set of its $l$ nearest neighbors, including itself. We consider four possible values $l \in \{3,4,5,6\}$ in the analysis. This specification assumes that the outcome of each city depends only on its connections to these neighbors and on the connections formed among them. 

\begin{figure}[!h]
   \centering
    \setlength{\abovecaptionskip}{-0.3cm}
    \includegraphics[width=0.9\linewidth]{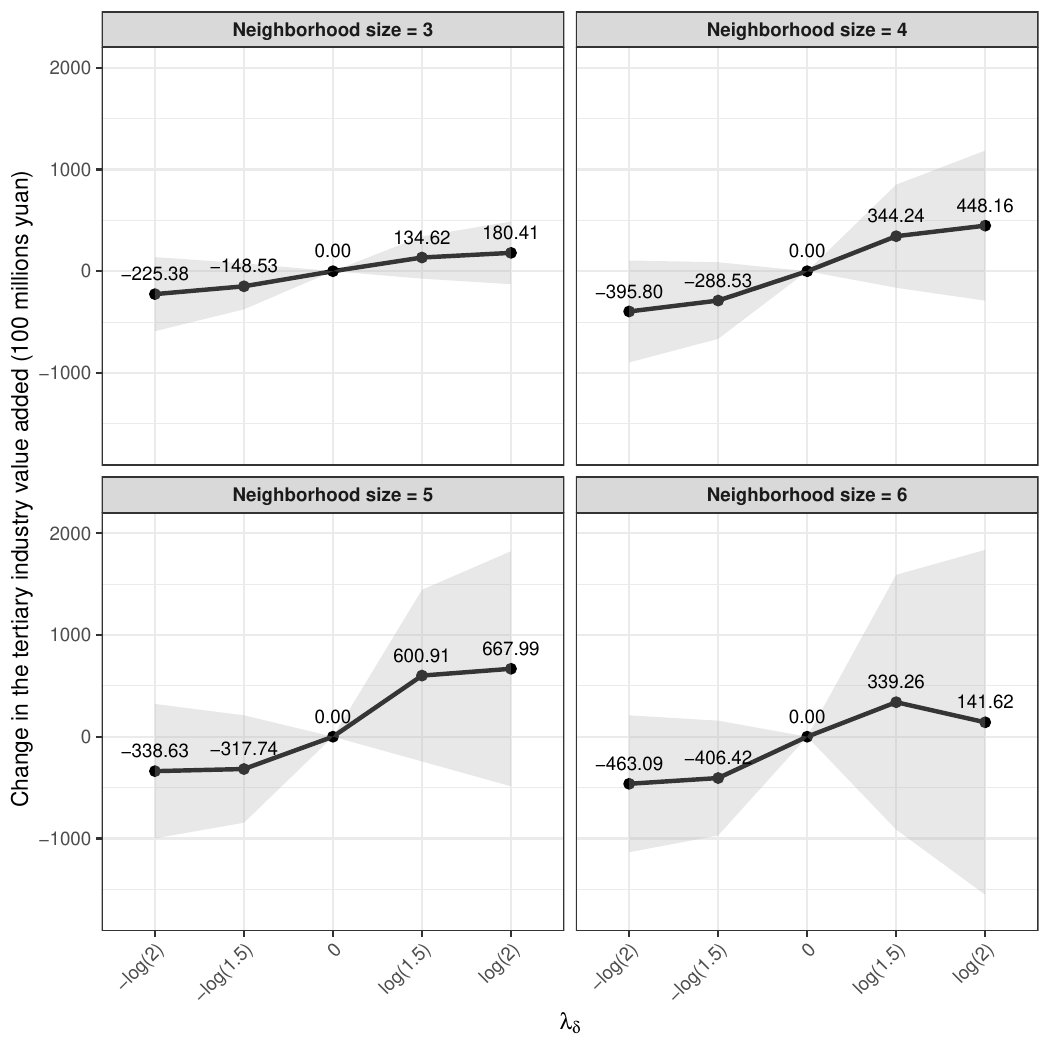}
    \caption{ The impact of China's railway network on the tertiary industry value added  under a 2-hour travel-time cutoff. Each panel displays results on for a fixed exclusion neighborhood size ($l \in {3, 4, 5, 6}$). The lines show the estimates of $\theta^\delta-\theta^0$ across varying levels of intervention intensity ($\lambda_\delta$), and the shaded regions indicate 95\% pointwise confidence intervals.}
    \label{fig:result_gdp}
\end{figure}

Figure~\ref{fig:result_gdp} reports the estimates based on the Hájek estimator $\hat{\theta}^\delta_2$. Across different intervention levels, the point estimates exhibit a generally increasing pattern, which is consistent with a potential positive association between improvements in local railway connectivity and tertiary industry value added. However, the corresponding variance estimates are sizable, leading to statistically insignificant effects at all intervention levels. We further observe that the estimated variance increases with the size of the exclusion neighborhood. This pattern is expected, as larger exclusion neighborhoods induce stronger cross-unit dependence and effectively reduce the amount of independent information available for estimation. Additional results based on alternative travel-time cutoffs are reported in the appendix, and the qualitative patterns remain similar. 

Several factors may account for these findings. First, the sample size may be too limited to yield precise estimates given the correlated structure of the data. Second, cities differ widely in their baseline conditions. Some may benefit substantially from improved connectivity, while others may see little impact due to their existing economic status, geographic location, or industrial composition. Third, the economic effects of infrastructure improvements often emerge over time. Since the analysis focuses on outcomes observed only one year after the intervention, it may not fully capture longer-term structural changes, especially in the service sector, where transitions tend to occur more slowly.

\section{Concluding remarks}
\label{sec::conclusion}
Edge interventions pose unique challenges, including complex interference patterns, limited overlap in treatment assignment, and the need for flexible modeling of treatment probabilities. We develop an approach for identifying and estimating the causal effects of edge interventions. Our approach leverages local interference, stochastic interventions, and ERGMs to address these challenges. The proposed IPW estimators are consistent and asymptotically Normal under appropriate conditions on the dependence structure of the edges.

The empirical analysis highlights both the feasibility and the intrinsic limitations of edge-level causal inference in network settings. While the point estimates suggest that increasing local rail connectivity may have a modest positive effect, the estimates are imprecise and not statistically distinguishable from zero. This result reflects the strong dependence induced by network structure and the resulting limits on effective sample size. Accordingly, the empirical results should be interpreted as demonstrating the implementation and inferential behavior of the framework in a realistic setting, rather than as delivering precise policy effect estimates.

Our work relates to the literature on group formation, which examines how forming groups with other units affects individual outcomes \citep[e.g.,][]{Xinran2019,basse2024randomization}. Group formation can be interpreted as establishing connections with other units. However, unlike our setting, connections formed by groups are mutually exclusive, and prior studies in this literature typically focus on randomized experiments under exposure-mapping assumptions about interference.

We defined a connection as an indicator for whether rail travel time between two cities is less than two hours. This dichotomization may obscure the interpretation of causal effects. The presence of a new connection may result from the construction of a new railway, an increase in train speed, or the development of an alternative route that shortens existing travel times. Future work could extend our framework to weighted edge interventions in order to distinguish among these mechanisms and provide more nuanced interpretations.

Another direction for extension is to consider temporal dynamics of edge connections. Rail networks evolve over time, and it is of interest to study how the dynamic evolution of rail transportation affects outcomes. Addressing this question requires extending the treatment from a single adjacency matrix to a sequence of adjacency matrices observed over time. We leave this important direction to future research.

\newpage

\bibliographystyle{Chicago}
\bibliography{network-ref}

@article{faber2014trade,
  title={Trade integration, market size, and industrialization: evidence from China's National Trunk Highway System},
  author={Faber, Benjamin},
  journal={Review of Economic Studies},
  volume={81},
  number={3},
  pages={1046--1070},
  year={2014},
  publisher={Oxford University Press}
}

@article{hainmueller2012entropy,
  title={Entropy balancing for causal effects: A multivariate reweighting method to produce balanced samples in observational studies},
  author={Hainmueller, Jens},
  journal={Political Analysis},
  volume={20},
  number={1},
  pages={25--46},
  year={2012},
  publisher={Cambridge University Press}
}

@article{sun2020displaying,
  title={Displaying things in common to encourage friendship formation: A large randomized field experiment},
  author={Sun, Tianshu and Taylor, Sean J},
  journal={Quantitative Marketing and Economics},
  volume={18},
  number={3},
  pages={237--271},
  year={2020},
  publisher={Springer}
}

@article{viviano2024policy,
  title={Policy design in experiments with unknown interference},
  author={Viviano, Davide and Rudder, Jess},
  journal={arXiv preprint arXiv:2011.08174
        
        
        
        
        
        
        
        },
  year={2024}
}

@article{gonzalez2016paving,
  title={Paving streets for the poor: Experimental analysis of infrastructure effects},
  author={Gonzalez-Navarro, Marco and Quintana-Domeque, Climent},
  journal={Review of Economics and Statistics},
  volume={98},
  number={2},
  pages={254--267},
  year={2016},
  publisher={The MIT Press}
}

@article{savje2021average,
  title={Average treatment effects in the presence of unknown interference},
  author={S{\"a}vje, Fredrik and Aronow, Peter and Hudgens, Michael},
  journal={Annals of Statistics},
  volume={49},
  number={2},
  pages={673},
  year={2021},
  publisher={NIH Public Access}
}

@article{kennedy2019nonparametric,
  title={Nonparametric causal effects based on incremental propensity score interventions},
  author={Kennedy, Edward H},
  journal={Journal of the American Statistical Association},
  volume={114},
  number={526},
  pages={645--656},
  year={2019},
  publisher={Taylor \& Francis}
}

@article{wu2024assessing,
  title={Assessing the causal effects of a stochastic intervention in time series data: are heat alerts effective in preventing deaths and hospitalizations?},
  author={Wu, Xiao and Weinberger, Kate R and Wellenius, Gregory A and Dominici, Francesca and Braun, Danielle},
  journal={Biostatistics},
  volume={25},
  number={1},
  pages={57--79},
  year={2024},
  publisher={Oxford University Press}
}

@article{papadogeorgou2022causal,
  title={Causal inference with spatio-temporal data: estimating the effects of airstrikes on insurgent violence in Iraq},
  author={Papadogeorgou, Georgia and Imai, Kosuke and Lyall, Jason and Li, Fan},
  journal={Journal of the Royal Statistical Society Series B: Statistical Methodology},
  volume={84},
  number={5},
  pages={1969--1999},
  year={2022},
  publisher={Oxford University Press}
}

@article{gao2025causal,
  author  = {Gao, Mengsi and Ding, Peng},
  title   = {Causal Inference in Network Experiments: Regression-Based Analysis and Design-Based Properties},
  journal = {arXiv preprint arXiv:2309.07476
        
        
        
        
        
        },
  year    = {2025}
}

@article{banerjee2020road,
  title={On the road: Access to transportation infrastructure and economic growth in China},
  author={Banerjee, Abhijit and Duflo, Esther and Qian, Nancy},
  journal={Journal of Development Economics},
  volume={145},
  pages={102442},
  year={2020},
  issn = {0304-3878},
  publisher={Elsevier}
}

@article{leung2022causal,
  title={Causal inference under approximate neighborhood interference},
  author={Leung, Michael P},
  journal={Econometrica},
  volume={90},
  number={1},
  pages={267--293},
  year={2022},
  publisher={Wiley Online Library}
}

@article{Frank1986,
author = {Ove Frank and David Strauss},
title = {Markov Graphs},
journal = {Journal of the American Statistical Association},
volume = {81},
number = {395},
pages = {832--842},
year = {1986},
publisher = {ASA Website},
doi = {10.1080/01621459.1986.10478342}
}

@article{Xinran2019,
author = {Li, Xinran and Ding, Peng and Lin, Qian and Yang, Dawei and Liu, Jun S.},
title = {Randomization Inference for Peer Effects},
journal = {Journal of the American Statistical Association},
volume = {114},
number = {528},
pages = {1651--1664},
year = {2019},
publisher = {ASA Website},
doi = {10.1080/01621459.2018.1512863 }
}

@article{abadie2020sampling,
  title={Sampling-based versus design-based uncertainty in regression analysis},
  author={Abadie, Alberto and Athey, Susan and Imbens, Guido W and Wooldridge, Jeffrey M},
  journal={Econometrica},
  volume={88},
  number={1},
  pages={265--296},
  year={2020},
  publisher={Wiley Online Library}
}

@book{imbens2015causal,
  title={Causal Inference for Statistics, Social, and Biomedical Sciences: An Introduction},
  author={Imbens, Guido W and Rubin, Donald B},
  year={2015},
  publisher={Cambridge university press}
}

@article{rosenbaum1983central,
  title={The central role of the propensity score in observational studies for causal effects},
  author={Rosenbaum, Paul R and Rubin, Donald B},
  journal={Biometrika},
  volume={70},
  number={1},
  pages={41--55},
  year={1983},
  publisher={Oxford University Press}
}

@article{kojevnikov2021limit,
  title={Limit theorems for network dependent random variables},
  author={Kojevnikov, Denis and Marmer, Vadim and Song, Kyungchul},
  journal={Journal of Econometrics},
  volume={222},
  number={2},
  pages={882--908},
  year={2021},
  publisher={Elsevier}
}

@article{snijders2006new,
  title={New specifications for exponential random graph models},
  author={Snijders, Tom AB and Pattison, Philippa E and Robins, Garry L and Handcock, Mark S},
  journal={Sociological Methodology},
  volume={36},
  number={1},
  pages={99--153},
  year={2006},
  publisher={Wiley Online Library}
}

@article{hunter2008goodness,
  title={Goodness of fit of social network models},
  author={Hunter, David R and Goodreau, Steven M and Handcock, Mark S},
  journal={Journal of the American Statistical Association},
  volume={103},
  number={481},
  pages={248--258},
  year={2008},
  publisher={Taylor \& Francis}
}

@article{schweinberger2015local,
  title={Local dependence in random graph models: characterization, properties and statistical inference},
  author={Schweinberger, Michael and Handcock, Mark S},
  journal={Journal of the Royal Statistical Society Series B: Statistical Methodology},
  volume={77},
  number={3},
  pages={647--676},
  year={2015},
  publisher={Oxford University Press}
}

@article{kojevnikov2021bootstrap,
  author  = {Kojevnikov, Denis},
  title   = {The Bootstrap for Network Dependent Processes},
  journal = {arXiv preprint arXiv:2101.12312
        
        
        
        },
  year    = {2021}
}

@article{ma2024distributional,
  title={The distributional impacts of transportation networks in {China}},
  author={Ma, Lin and Tang, Yang},
  journal={Journal of International Economics},
  volume={148},
  pages={103873},
  year={2024},
  publisher={Elsevier}
}

@article{you2018relief,
  title={Relief degree of land surface dataset of China (1 km)},
  author={You, Zhen and Feng, ZM and Yang, YZ and others},
  journal={Journal of Global Change Data and Discovery},
  volume={2},
  number={2},
  pages={151--155},
  year={2018}
}

@article{aronow2017estimating,
  title={Estimating average causal effects under general interference, with application to a social network experiment},
journal={The Annals of Applied Statistics},
  author={Aronow, Peter M and Samii, Cyrus},
  year={2017},
number = {4},
 pages = {1912--1947},
volume = {11}
}

@article{savje2024causal,
  title={Causal inference with misspecified exposure mappings: separating definitions and assumptions},
  author={S{\"a}vje, Fredrik},
  journal={Biometrika},
  volume={111},
  number={1},
  pages={1--15},
  year={2024},
  publisher={Oxford University Press}
}

@article{holland1981exponential,
  title={An exponential family of probability distributions for directed graphs},
  author={Holland, Paul W and Leinhardt, Samuel},
  journal={Journal of the American Statistical Association},
  volume={76},
  number={373},
  pages={33--50},
  year={1981},
  publisher={Taylor \& Francis}
}

@article{owusu2025randomization,
  author  = {Owusu, Julius},
  title   = {Randomization Inference of Heterogeneous Treatment Effects under Network Interference},
  journal = {arXiv preprint arXiv:2308.00202},
  year    = {2025},
  url     = {https://arxiv.org/abs/2308.00202}
}

@article{zhou2024estimating,
  title={Estimating heterogeneous treatment effects for spatio-temporal causal inference: How economic assistance moderates the effects of airstrikes on insurgent violence},
  author={Zhou, Lingxiao and Imai, Kosuke and Lyall, Jason and Papadogeorgou, Georgia},
  journal={arXiv preprint arXiv:2412.15128},
  year={2024}
}

@article{shook2022power,
  title={Power and sample size for observational studies of point exposure effects},
  author={Shook-Sa, Bonnie E and Hudgens, Michael G},
  journal={Biometrics},
  volume={78},
  number={1},
  pages={388--398},
  year={2022},
  publisher={Wiley Online Library}
}

@article{schweinberger2020concentration,
  title={Concentration and consistency results for canonical and curved exponential-family models of random graphs},
  author={Schweinberger, Michael and Stewart, Jonathan},
  journal={The Annals of Statistics},
  volume={48},
  number={1},
  pages={374--396},
  year={2020},
  publisher={JSTOR}
}

@article{schweinberger2020exponential,
  title={Exponential-family models of random graphs},
  author={Schweinberger, Michael and Krivitsky, Pavel N and Butts, Carter T and Stewart, Jonathan R},
  journal={Statistical Science},
  volume={35},
  number={4},
  pages={627--662},
  year={2020},
  publisher={JSTOR}
}

@article{wang2022china,
  title={How China's high-speed rail promote local economy: New evidence from county-level panel data},
  author={Wang, Yao and Dong, Weijia},
  journal={{International Review of Economics and Finance}},
  volume={80},
  pages={67--81},
  year={2022},
  publisher={Elsevier}
}

@article{wang2020impact,
  title={The impact of high-speed rails on urban economy: An investigation using night lighting data of Chinese cities},
  author={Wang, Chunyang and Meng, Weidong and Hou, Xinshuo},
  journal={Research in Transportation Economics},
  volume={80},
  pages={100819},
  year={2020},
  publisher={Elsevier}
}

@article{basse2024randomization,
  title={Randomization tests for peer effects in group formation experiments},
  author={Basse, Guillaume and Ding, Peng and Feller, Avi and Toulis, Panos},
  journal={Econometrica},
  volume={92},
  number={2},
  pages={567--590},
  year={2024},
  publisher={Wiley Online Library}
}

@article{sobel2006randomized,
  title={What do randomized studies of housing mobility demonstrate? Causal inference in the face of interference},
  author={Sobel, Michael E},
  journal={Journal of the American Statistical Association},
  volume={101},
  number={476},
  pages={1398--1407},
  year={2006},
  publisher={Taylor \& Francis}
}

@article{hudgens2008toward,
  title={Toward causal inference with interference},
  author={Hudgens, Michael G and Halloran, M Elizabeth},
  journal={Journal of the American Statistical Association},
  volume={103},
  number={482},
  pages={832--842},
  year={2008},
  publisher={Taylor \& Francis}
}

@article{diaz2020causal,
  title={Causal mediation analysis for stochastic interventions},
  author={D{\'\i}az, Iv{\'a}n and Hejazi, Nima S},
  journal={Journal of the Royal Statistical Society Series B: Statistical Methodology},
  volume={82},
  number={3},
  pages={661--683},
  year={2020},
  publisher={Oxford University Press}
}

@article{diaz2023nonparametric,
  title={Nonparametric causal effects based on longitudinal modified treatment policies},
  author={D{\'\i}az, Iv{\'a}n and Williams, Nicholas and Hoffman, Katherine L and Schenck, Edward J},
  journal={Journal of the American Statistical Association},
  volume={118},
  number={542},
  pages={846--857},
  year={2023},
  publisher={Taylor \& Francis}
}

@article{manski1993identification,
  title={Identification of endogenous social effects: The reflection problem},
  author={Manski, Charles F},
  journal={The Review of Economic Studies},
  volume={60},
  number={3},
  pages={531--542},
  year={1993},
  publisher={Wiley-Blackwell}
}

@article{li2019propensity,
  title={Propensity score weighting for causal inference with multiple treatments},
  author={Li, Fan and Li, Fan},
  journal={Statistical Methods in Medical Research},
  volume={28},
  number={3},
  pages={785--804},
  year={2019},
  publisher={SAGE Publications Sage UK: London, England},
  doi={10.1177/0962280218799713},
  url={https://doi.org/10.1177/0962280218799713}
}

@book{ding2024first,
  title={A First Course in Causal Inference},
  author={Ding, Peng},
  year={2024},
  publisher={Chapman and Hall/CRC}
}

@article{papadogeorgou2025causal,
  author  = {Papadogeorgou, Georgia and Song, Zhaoyan and Imbens, Guido and Mealli, Fabrizia},
  title   = {Causal Inference when Intervention Units and Outcome Units Differ},
  journal = {arXiv preprint arXiv:2507.20231
        
        
        
        
        
        
        
        
        
        
        
        
        
        },
  year    = {2025},
  url     = {https://arxiv.org/abs/2507.20231}
}

@article{kennedy2017non,
  title={Nonparametric methods for doubly robust estimation of continuous treatment effects},
  author={Kennedy, Edward H and Ma, Zongming and McHugh, Matthew D and Small, Dylan S},
  journal={Journal of the Royal Statistical Society Series B: Statistical Methodology},
  volume={79},
  number={4},
  pages={1229--1245},
  year={2017},
  publisher={Oxford University Press}
}

@article{jiang2023instrumental,
  title={An instrumental variable method for point processes: generalized Wald estimation based on deconvolution},
  author={Jiang, Zhichao and Chen, Shizhe and Ding, Peng},
  journal={Biometrika},
  volume={110},
  number={4},
  pages={989--1008},
  year={2023},
  publisher={Oxford University Press}
}

@article{tan2025causal,
  title={Causal effect of functional treatment},
  author={Tan, Ruoxu and Huang, Wei and Zhang, Zheng and Yin, Guosheng},
  journal={Journal of Machine Learning Research},
  volume={26},
  number={91},
  pages={1--39},
  year={2025}
}

@article{snijders2002markov,
  title={Markov chain Monte Carlo estimation of exponential random graph models},
  author={Snijders, Tom AB and others},
  journal={Journal of Social Structure},
  volume={3},
  number={2},
  pages={1--40},
  year={2002}
}

@article{hunter2008ergm,
  title={ergm: A Package to Fit, Simulate and Diagnose Exponential-Family Models for Networks},
  author={Hunter, David R and Handcock, Mark S and Butts, Carter T and Goodreau, Steven M and Morris, Martina},
  journal={Journal of Statistical Software},
  volume={24},
  number={3},
  pages={1--29},
  year={2008}
}

@article{imbens2000role,
  title={The role of the propensity score in estimating dose-response functions},
  author={Imbens, Guido W},
  journal={Biometrika},
  volume={87},
  number={3},
  pages={706--710},
  year={2000},
  publisher={Oxford University Press}
}

@article{imai2004causal,
  title={Causal inference with general treatment regimes: Generalizing the propensity score},
  author={Imai, Kosuke and Van Dyk, David A},
  journal={Journal of the American Statistical Association},
  volume={99},
  number={467},
  pages={854--866},
  year={2004},
  publisher={Taylor \& Francis}
}

@article{papadogeorgou2019causal,
  title={Causal inference with interfering units for cluster and population level treatment allocation programs},
  author={Papadogeorgou, Georgia and Mealli, Fabrizia and Zigler, Corwin M},
  journal={Biometrics},
  volume={75},
  number={3},
  pages={778--787},
  year={2019},
  publisher={Wiley Online Library}
}

@article{barkley2020causal,
author = {Brian G. Barkley and Michael G. Hudgens and John D. Clemens and Mohammad Ali and Michael E. Emch},
title = {{Causal inference from observational studies with clustered interference, with application to a cholera vaccine study}},
volume = {14},
journal = {The Annals of Applied Statistics},
number = {3},
publisher = {Institute of Mathematical Statistics},
pages = {1432 -- 1448},
year = {2020}
}

@article{manski2013identification,
  title={Identification of treatment response with social interactions},
  author={Manski, Charles F},
  journal={The Econometrics Journal},
  volume={16},
  number={1},
  pages={S1--S23},
  year={2013},
  publisher={Oxford University Press Oxford, UK}
}

@article{forastiere2021identification,
  title={Identification and estimation of treatment and interference effects in observational studies on networks},
  author={Forastiere, Laura and Airoldi, Edoardo M and Mealli, Fabrizia},
  journal={Journal of the American Statistical Association},
  volume={116},
  number={534},
  pages={901--918},
  year={2021},
  publisher={Taylor \& Francis}
}

@article{wang2025design,
  title={Design-based inference for spatial experiments under unknown interference},
  author={Wang, Ye and Samii, Cyrus and Chang, Haoge and Aronow, PM},
  journal={The Annals of Applied Statistics},
  volume={19},
  number={1},
  pages={744--768},
  year={2025},
  publisher={Institute of Mathematical Statistics}
}

@article{lee2025efficient,
  title={Efficient nonparametric estimation of stochastic policy effects with clustered interference},
  author={Lee, Chanhwa and Zeng, Donglin and Hudgens, Michael G},
  journal={Journal of the American Statistical Association},
  volume={120},
  number={549},
  pages={382--394},
  year={2025},
  publisher={Taylor \& Francis}
}

@article{zhang2021covariate,
  title={Covariate balancing functional propensity score for functional treatments in cross-sectional observational studies},
  author={Zhang, Xiaoke and Xue, Wu and Wang, Qiyue},
  journal={Computational Statistics and Data Analysis},
  volume={163},
  pages={107303},
  year={2021},
  publisher={Elsevier}
}

@article{sussman2017elements,
  author  = {Sussman, Daniel L. and Airoldi, Edoardo M.},
  title   = {Elements of Estimation Theory for Causal Effects in the Presence of Network Interference},
  journal = {arXiv preprint arXiv:1702.03578
        
        },
  year    = {2017},
  url     = {https://arxiv.org/abs/1702.03578}
}

@article{jagadeesan2020designs,
  title={Designs for estimating the treatment effect in networks with interference},
  author={Jagadeesan, Ravi and Pillai, Natesh S and Volfovsky, Alexander},
journal={Annals of Statistics},
volume = {48},
number = {2},
 pages={679--712},
  year={2020}
}

@inproceedings{awan2020almost,
   title = 	 {Almost-Matching-Exactly for Treatment Effect Estimation under Network Interference},
  author =       {Awan, Usaid and Morucci, Marco and Orlandi, Vittorio and Roy, Sudeepa and Rudin, Cynthia and Volfovsky, Alexander},
  booktitle = 	 {Proceedings of the Twenty Third International Conference on Artificial Intelligence and Statistics},
  pages = 	 {3252--3262},
  year = 	 {2020},
  volume = 	 {108},
  publisher =    {PMLR}
}

@article{forastiere2022estimating,
  title={Estimating causal effects under network interference with bayesian generalized propensity scores},
  author={Forastiere, Laura and Mealli, Fabrizia and Wu, Albert and Airoldi, Edoardo M},
  journal={Journal of Machine Learning Research},
  volume={23},
  number={289},
  pages={1--61},
  year={2022}
}

@article{karwa2018systematic,
  author  = {Karwa, Vishesh and Airoldi, Edoardo M.},
  title   = {A Systematic Investigation of Classical Causal Inference Strategies under Mis-specification due to Network Interference},
  journal = {arXiv preprint arXiv:1810.08259
        
        
        
        
        
        
        
        
        
        
        
        },
  year    = {2018}
}

@article{belloni2025neighborhood,
      title={Neighborhood Adaptive Estimators for Causal Inference under Network Interference}, 
      author={Alexandre Belloni and Fei Fang and Alexander Volfovsky},
      year={2025},
      journal = {arXiv preprint arXiv:2212.03683
        
        
        
         }
}

@article{jorritsma2009substitution, 
  title={Substitution opportunities of High Speed Train for air transport},
  author={Jorritsma, Peter},
  journal={Transport Business},
  volume={7},
  pages={121--128},
  year={2009}
}

@article{kroes2019substitution,
  title={Substitution from air to high-speed rail: the case of Amsterdam airport},
  author={Kroes, Eric and Savelberg, Fons},
  journal={Transportation Research Record},
  volume={2673},
  number={5},
  pages={166--174},
  year={2019},
  publisher={SAGE Publications Sage CA: Los Angeles, CA}
}

@article{he2016tolerance,
  title={Tolerance threshold of commuting time: evidence from Kunming, China},
  author={He, Mingwei and Zhao, Shengchuan and He, Min},
  journal={Journal of Transport Geography},
  volume={57},
  pages={1--7},
  year={2016},
  publisher={Elsevier}
}

@article{de2025makes,
  title={What makes a commute enjoyable: A duration close to the ideal, or far below the maximum tolerable?},
  author={De Vos, Jonas},
  journal={Travel Behaviour and Society},
  volume={41},
  pages={101095},
  year={2025},
  publisher={Elsevier}
}

@InProceedings{Toulis2013Estimation,
  title = 	 {Estimation of Causal Peer Influence Effects},
  author = 	 {Toulis, Panos and Kao, Edward},
  booktitle = 	 {Proceedings of the 30th International Conference on Machine Learning},
  pages = 	 {1489--1497},
  year = 	 {2013},
  editor = 	 {Dasgupta, Sanjoy and McAllester, David},
  volume = 	 {28},
  series = 	 {Proceedings of Machine Learning Research},
  address = 	 {Atlanta, Georgia, USA},
  publisher =    {PMLR}
  
}

@article{Bargagli2025HETEROGENEOUS,
journal = {The Annals of Applied Statistics},
issn = {1932-6157},
number = {1},
pmid = {40642103},
pmcid = {PMC12245184
        
        
        
        
        
         },
address = {United States},
title = {HETEROGENEOUS TREATMENT AND SPILLOVER EFFECTS UNDER CLUSTERED NETWORK INTERFERENCE},
volume = {19},
author = {Bargagli-Stoffi, Falco J and Tortú, Costanza and Forastiere, Laura},
pages = {28--55},
date = {2025},
year = {2025}
}

@article{basse2018analyzing,
  title={Analyzing two-stage experiments in the presence of interference},
  author={Basse, Guillaume and Feller, Avi},
  journal={Journal of the American Statistical Association},
  volume={113},
  number={521},
  pages={41--55},
  year={2018},
  publisher={Taylor \& Francis}
}

@article{geyer1992constrained,
  title={Constrained Monte Carlo maximum likelihood for dependent data},
  author={Geyer, Charles J and Thompson, Elizabeth A},
  journal={Journal of the Royal Statistical Society: Series B (Methodological)},
  volume={54},
  number={3},
  pages={657--683},
  year={1992},
  publisher={Wiley Online Library}
}

\newpage
\appendix
\pagenumbering{arabic} 
\renewcommand*{\thepage}{S\arabic{page}}

\setcounter{equation}{0}
\setcounter{figure}{0}
\setcounter{theorem}{0}
\setcounter{lemma}{0}
\setcounter{section}{0}
\setcounter{corollary}{0}
\setcounter{example}{0}
\setcounter{definition}{0}
\renewcommand {\theequation} {S\arabic{equation}}
\renewcommand {\thefigure} {S\arabic{figure}}
\renewcommand {\thetheorem} {S\arabic{theorem}}
\renewcommand {\thelemma} {S\arabic{lemma}}
\renewcommand {\thesection} {S\arabic{section}}
\renewcommand {\thecorollary} {S\arabic{corollary}}
\renewcommand {\theexample} {S\arabic{example}}
\renewcommand {\thedefinition} {S\arabic{definition}}
\begin{center}
  \LARGE {\bf Supplementary Material}
\end{center}

Section~\ref{Appendix A} provides proofs of the results. 

Section~\ref{app::sim} reports additional simulation results.

Section~\ref{app::application} presents more details on the empirical analysis.
  
\section{Proofs}\label{Appendix A}
\subsection{Proof of Theorem~\ref{th::identification}}
For each unit $i$, we have
\begin{eqnarray*}
\E\left\{\frac{\pr(\bm{A}^{\delta}_{i})}{\pr(\bm{A}_{i})}Y_i\right\}
&=&\E\left\{\sum_{\bm{a}_{i}\in \mathcal{A}_i} \frac{\pr(\bm{A}^{\delta}_{i}=\bm{a}_{i})}{\pr(\bm{A}_{i}=\bm{a}_{i})}Y_i(\bm{a}_{i}) \bm{1}(\bm{A}_{i}=\bm{a}_{i}) \right\}\\
&=&\sum_{\bm{a}_{i}\in \mathcal{A}_i} \frac{\pr(\bm{A}^{\delta}_{i}=\bm{a}_{i})}{\pr(\bm{A}_{i}=\bm{a}_{i})}Y_i(\bm{a}_{i}) \pr(\bm{A}_i=\bm{a}_{i}) \\
&=&\theta_i^{\delta}.
\end{eqnarray*}
This leads to the identification formula in Theorem~\ref{th::identification}. \QEDB

\subsection{Proof of Theorem \ref{Th::Consistency of theta}}
It suffices to show that $\var\left(\hat{\theta}^{\delta}\right)\rightarrow 0$.
Define
\begin{eqnarray*}
    \tilde{Y}_i&=& \frac{\exp\{e(\bm{A}_i)\lambda_\delta\}}{\E_{\hat{\eta}}\left\{ \exp\{e(\bm{A}_i)\lambda_\delta\}\right\}}Y_i(\bm{A}_i).
\end{eqnarray*}
Then $\hat{\theta}_1^\delta$ can be written as $ n^{-1} \sum_{i=1}^n \tilde{Y}_i$.
By definition, if $j\notin \mathcal{R}_i$, then we have $(\bm{A}_i,\bm{A}^\delta_i) \ind (\bm{A}_j,\bm{A}^\delta_j)$, which implies $\tilde{Y}_i\ind \tilde{Y}_j$. 
As a result, we have
\begin{eqnarray*}
     \var\left(\hat{\theta}^{\delta}_1\right)&=&\frac{1}{n^2} \sum_{i=1}^n \sum_{j=1}^n\cov(\tilde{Y}_i,\tilde{Y}_j)\\
    &=&\frac{1}{n^2} \sum_{i=1}^n \sum_{j=1}^n \bone(j\in \mathcal{R}_i)\cov\left(\tilde{Y}_i,\tilde{Y}_j\right)\\
    &\leq& \frac{C}{n^2} \sum_{i=1}^n \sum_{j=1}^n \bone(j\in \mathcal{R}_i)\\
    &=& \frac{C}{n^2}\sum_{i=1}^n |\mathcal{R}_i|
\end{eqnarray*}
for some constant $C>0$. From the condition in the theorem, we obtain that $\var\left(\hat{\theta}^{\delta}\right)$ converges to zero. \QEDB

\subsection{Proof of Theorem \ref{Th::CLT}}

We apply Stein's method for the proof. We first introduce some definitions and lemmas for the Stein's method.
\begin{definition}
Let $W$ and $Z$ be two random variables. Let $\mathcal{G}$ denote the set of all Lipschitz functions $g : \mathbb{R} \to \mathbb{R}$.
\begin{enumerate}[(a)]
  \item The Kolmogorov–Smirnov distance is defined as:
  \begin{equation*}
      \mathrm{Kolm}(W, Z) = \sup_{x \in \mathbb{R}} \left| \pr(W \leq x) - \pr(Z \leq x) \right|.
  \end{equation*}

  \item The Wasserstein distance is defined as:
  \begin{equation*}
      \mathrm{Wass}(W, Z) = \sup_{g \in \mathcal{G}} \left| \mathbb{E}[g(W)] - \mathbb{E}[g(Z)] \right|.
  \end{equation*}
\end{enumerate}
\end{definition}
\begin{lemma}
\label{lem::1}
Suppose that $Z$ follows the standard Normal distribution. Then,
\begin{equation*}
    \mathrm{Kolm}(W, Z) \leq 2 \sqrt{ \mathrm{Wass}(W, Z)/\sqrt{2\pi}}.
    \label{eq:kolm-wass-bound}
\end{equation*}
\end{lemma}

 \begin{lemma}
 \label{lem::2}
 Suppose that $Z$ follows the standard Normal distribution. Then,
 \begin{eqnarray*}
     \mathrm{Wass}(W, Z)&=& \sup_{g \in \mathcal{G}} \left| \mathbb{E}[g(W)] - \mathbb{E}[g(Z)] \right| 
\ \leq\ \sup_{f \in \mathcal{F}} \left| \mathbb{E}\left[ f'(W) - W f(W) \right] \right|,
 \end{eqnarray*}
where
$$
\mathcal{F}\ =\ \left\{ f \in \mathcal{F} : \|f\|_\infty \leq 1, \|f'\|_\infty \leq \sqrt{\frac{2}{\pi}}, \|f''\|_\infty \leq 2 \right\}.
$$
 \end{lemma}
The proofs of these lemmas are standard and thus omitted.

\medskip

We then prove Theorem~\ref{Th::CLT}.
Define 
\begin{eqnarray*}
   W_i&=&  \frac{\exp\{e(\bm{A}_i)\lambda_\delta\}}{\E_{\hat{\eta}}\left\{ \exp\{e(\bm{A}_i)\lambda_\delta\}\right\}}Y_i(\bm{A}_i) - \theta^\delta_i, \quad \sigma^2 \ =\ \var\left( \sum_{i=1}^n W_i \right),\quad
   W\ = \ \frac{ \sum_{i=1}^n W_i }{\sigma}.
\end{eqnarray*}
We can write
\begin{eqnarray*}
    \frac{\hat{\theta}^\delta_1-\theta_\delta}{\sqrt{\var(\hat{\theta}^\delta_1)}}&=& W.
\end{eqnarray*}

We derive the bounds on $\mathrm{Wass}(W, Z) $ based on Lemma~\ref{lem::2}.
Define 
\begin{eqnarray*}
    W_{-i}&=& \frac{ \sum_{j=1}^n \bm{1}(j \notin \mathcal{R}_i)W_j }{\sigma}.
\end{eqnarray*}
From the proof of Theorem~\ref{Th::Consistency of theta}, we have $W_i\ind W_{-i}$.
For any function $f \in \mathcal{F}$, we have 
\begin{eqnarray*}
        \E\{Wf(W)\} &=&\frac{1}{\sigma}\sum_{i=1}^n\E\{W_if(W)\}
\ =\ \frac{1}{\sigma}\sum_{i=1}^n\mathbb{E}\left[W_i\{f(W)-f(W_{-i})\}\right],
    \end{eqnarray*}
where the second equality comes from $\E\{W_if(W_{-i})\}=\E(W_i)\E\{f(W_{-i})\}=0$.
We can further decompose $\E\{Wf(W)\}$ as $\E\{Wf(W)\}=B_1 + B_2$, where
    \begin{eqnarray*}
  B_1&=&      \frac{1}{\sigma} \sum_{i=1}^n \E\left[W_i \left\{f(W) - f(W_{-i}) - (W - W_{-i}) f'(W)\right\}\right],\\
  B_2 &=& \frac{1}{\sigma} \sum_{i=1}^n \E\left\{W_i (W - W_{-i}) f'(W)\right\}.
    \end{eqnarray*}
Applying Taylor expansion for $B_1$, we obtain
\begin{eqnarray*}
B_1  &\leq&
 \frac1{2\sigma}|f^{\prime\prime}|_{\infty}\sum_{i=1}^n\E\big|W_i(W-W_{-i})^2\big| 
  \ \leq\ \frac{1}{\sigma^3}\sum_{i=1}^n\E\left|W_i\left(\sum_{j\in \mathcal{R}_i}W_j\right)^2\right|.
\end{eqnarray*}
For $B_2$, we can write  $B_2=\E\{f^{\prime}(W)T\}$, where
\begin{eqnarray*}
    T&=& \frac{1}{\sigma}\sum_i W_i(W-W_{-i}).
\end{eqnarray*}
As a result, we have
\begin{eqnarray*}
    B_2-f^\prime(W)&\leq&|f^{\prime}|_\infty\cdot\mathbb{E}|T-1| 
 \ \leq\ \sqrt{\frac{2}{\pi}}\sqrt{\mathbb{E}(T-1)^2}
\ =\ \sqrt{\frac{2}{\pi}}\sqrt{\mathrm{Var}(T)},
\end{eqnarray*}
where the last equality follows from
\begin{eqnarray*}
    \E(T)&=&\frac{1}{\sigma}\sum_{i=1}^n\E\left(W W_i\right)\ =\ \E\left(W^2\right)\ =\ 1.
\end{eqnarray*}
Therefore, 
\begin{eqnarray}
\nonumber &&\left|\mathbb{E}\{W f(W)\} - \mathbb{E}\{f'(W)\}\right| \\
\nonumber&\leq& \frac{1}{\sigma^3}\sum_{i=1}^n\E\left|W_i\left(\sum_{j\in \mathcal{R}_i}W_j\right)^2\right|+\sqrt{\frac{2}{\pi}}\sqrt{\mathrm{Var}(T)} \\
\nonumber&=&\frac{1}{\sigma^3}\sum_{i=1}^n\sum_{j,k\in \mathcal{R}_i}\E\left|W_iW_jW_k\right|+
\sqrt{\frac{2}{\pi}} \cdot \sqrt{\var\left(\frac{1}{\sigma^2}\sum_{i=1}^n\sum_{j\in \mathcal{R}_i}W_iW_j\right)} \\
\label{eqn::steineqn}&\leq& \frac{1}{3\sigma^3} \sum_i \sum_{j,k \in \mathcal{R}_i} \left(\mathbb{E}|W_i|^3 + \mathbb{E}|W_j|^3 + \mathbb{E}|W_k|^3\right) +\sqrt{\frac{2}{\pi}} \cdot \sqrt{\var\left(\frac{1}{\sigma^2}\sum_{i=1}^n\sum_{j\in \mathcal{R}_i}W_iW_j\right)},
\end{eqnarray}
where the last equality follows from the AM-GM inequality.
Denote $D = \max_i|\mathcal{R}_i|$.
For the first term on the right hand side of~\eqref{eqn::steineqn}, we have 
\begin{eqnarray}
  \nonumber  && \sum_{i=1}^n \sum_{j,k \in \mathcal{R}_i} \left(\mathbb{E}|W_i|^3 + \mathbb{E}|W_j|^3 + \mathbb{E}|W_k|^3\right)\\
 \nonumber   &=& \sum_{i=1}^n |\mathcal{R}_i|^2|W_i|^3+ 2\sum_{i=1}^n \sum_{j\in \mathcal{R}_i} |\mathcal{R}_i| |W_j|^3\\
  \nonumber  &\leq &  D^2\sum_{i=1}^n |W_i|^3+ 2D\sum_{i=1}^n \sum_{j\in \mathcal{R}_i}  |W_j|^3\\
  \nonumber   &=&  D^2\sum_{i=1}^n |W_i|^3+ 2D\sum_{j=1}^n\sum_{i=1}^n \bm{1}(j\in \mathcal{R}_i) |W_j|^3\\
 \nonumber   &= &D^2\sum_{i=1}^n |W_i|^3+ 2D\sum_{j=1}^n\sum_{i=1}^n \bm{1}(i\in \mathcal{R}_j) |W_j|^3\\
 \label{eqn::normality1}   &\leq&3D^2\sum_{i=1}^n |W_i|^3.
\end{eqnarray}
where the last equality follows from the equivalence between $i \in \mathcal{R}_j$ and $j \in \mathcal{R}_i$.

For the second term on the right hand side of~\eqref{eqn::steineqn}, we have
\begin{eqnarray*}
    \var\left(\sum_{i=1}^n\sum_{j\in \mathcal{R}_i}W_iW_j\right)
&\leq& \sum_{i=1}^n\sum_{j\in \mathcal{R}_i}\sum_{k=1}^n\sum_{l\in \mathcal{R}_k} |\cov(W_iW_j,W_kW_l)|.
\end{eqnarray*}
If $k,l \notin \mathcal{R}_i\cap \mathcal{R}_j$, then $W_iW_j$ is independent of $W_kW_l$. 
Because $|\mathcal{R}_i\cap \mathcal{R}_j|\leq 2D$ and each unit in $\mathcal{R}_i\cap \mathcal{R}_j$ has at most $D$ neighbors, we have 
\begin{eqnarray*}
    \sum_{k=1}^n\sum_{l\in \mathcal{R}_k} |\cov(W_iW_j,W_kW_l)|&\leq&  2D^2 \max_{k\in \mathcal{N},l\in \mathcal{R}_k}|\cov(W_iW_j,W_kW_l)|\\
    &=& 2D^2 \frac{\max_{k\in \mathcal{N},l\in \mathcal{R}_k}|\cov(W_iW_j,W_kW_l)|}{\var(W_iW_j)}\var(W_iW_j)\\
    &\leq &2CD^2 \var(W_iW_j),
\end{eqnarray*}
where the constant 
$C$ is chosen such that 
$$\frac{\max_{k\in \mathcal{N},l\in \mathcal{R}_k}|\cov(W_iW_j,W_kW_l)|}{\var(W_iW_j)}\leq C.$$
Therefore, we have 
\begin{eqnarray}
 \nonumber   \var\left(\sum_{i=1}^n\sum_{j\in \mathcal{R}_i}W_iW_j\right)&\leq& 2CD^2 \sum_{i=1}^n\sum_{j\in \mathcal{R}_i} \var(W_iW_j)\\
\nonumber&\leq&2CD^2 \sum_{i=1}^n\sum_{j\in \mathcal{R}_i} \E(W_i^2W_j^2)\\
\nonumber&\leq& CD^2 \sum_{i=1}^n\sum_{j\in \mathcal{R}_i} \left\{\E(W_i^4) + \E(W_j^4)\right\}\\
\nonumber&\leq& CD^3\sum_{i=1}^n\E(W_i^4)+CD^2\sum_{i=1}^n \bm{1}(i\in \mathcal{R}_j)\E(W_j^4)\\
\label{eqn::normality2}&\leq& 2CD^3\sum_{i=1}^n\E(W_i^4).
\end{eqnarray}
Plugging~\eqref{eqn::normality1}~and~\eqref{eqn::normality2} into~\eqref{eqn::steineqn}, we obtain
\begin{eqnarray}
\label{eqn::steineqn2}
    \left|\mathbb{E}\{W f(W)\} - \mathbb{E}\{f'(W)\}\right| &\leq & \frac{D^2}{\sigma^3} \sum_{i=1}^n \mathbb{E}|W_i|^3+\frac{2C}{\sqrt{\pi}\sigma^2} \cdot \sqrt{D^3\sum_{i=1}^n\mathbb{E}(W_i^4)}.
\end{eqnarray}
Because the outcome is bounded, we have 
\begin{eqnarray}
\label{eqn::rate1}
    \sum_{i=1}^n \mathbb{E}|W_i|^3\ =\ O_\pr(n), \quad \sum_{i=1}^n \mathbb{E}(W_i^4) \ =\ O_\pr(n).
\end{eqnarray}
By definition, $ \var(\hat{\theta}^\delta_1)=\sigma^2/n^2 $. Therefore, the condition on $\max_i|\mathcal{R}_i|$ implies
\begin{eqnarray}
\label{eqn::rate2}
 \frac{n D^2}{\sigma^3} \ = \ o_\pr(1),\quad \frac{ n^{1/2} D^{3/2}}{\sigma^2}\ = \ o_\pr(1). 
\end{eqnarray}
Plugging~\eqref{eqn::rate1}~and~\eqref{eqn::rate2} into~\eqref{eqn::steineqn2}, we obtain that $|\mathbb{E}\{W f(W)\} - \mathbb{E}\{f'(W)\}|=o_\pr(1)$. The asymptotic Normality of $\hat{\theta}^\delta_1$ then follows from Lemmas~\ref{lem::1}~and~\ref{lem::2}. \QEDB

\section{Additional simulation results}
\label{app::sim}

\subsection{Simulation results for  $\hat{\theta}_1^\delta$ under Bernoulli random graph model}
\begin{figure}[htp]
    \centering
    \setlength{\abovecaptionskip}{-0.3cm}
    \includegraphics[width=16cm]{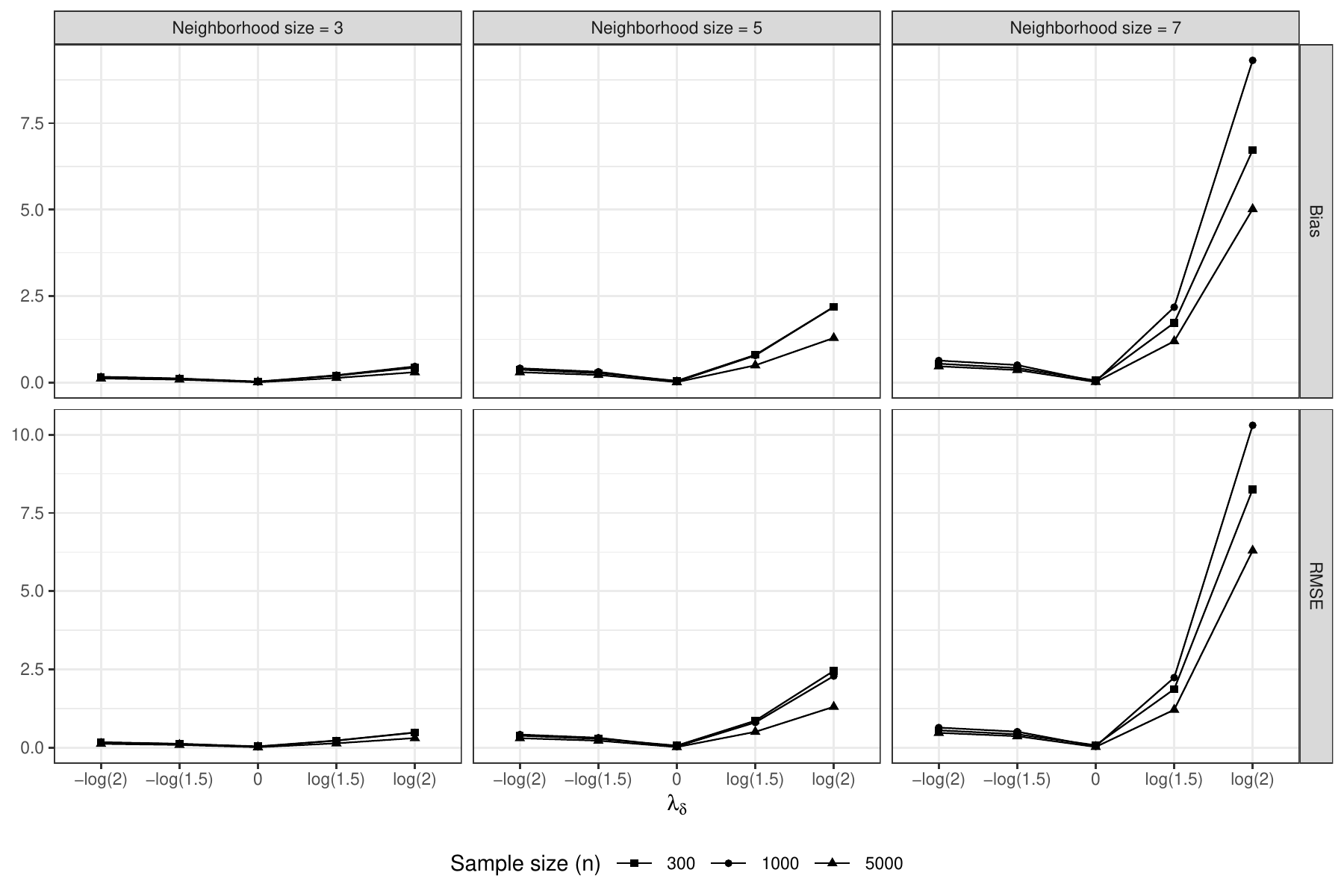}
\caption{Bias and RMSE of  \(\hat{\theta}^{\delta}_{1}\). The first row displays bias, and the second row displays RMSE. Each column represents a different exclusion neighborhood size $l \in \{3, 5, 7\}$.}
    \label{fig:ht_sim_bias_rmse}
\end{figure}

 Figure \ref{fig:ht_sim_bias_rmse} shows the bias and RMSE of $\hat{\theta}_1^\delta$ under the same data-generating process used in the main text. Compared to the H\'ajek estimator $\hat{\theta}_2^\delta$,  $\hat{\theta}_1^\delta$ exhibits larger bias and RMSE as the neighborhood size and intervention intensity increase. This pattern is particularly pronounced when both quantities are large. Nonetheless, both bias and RMSE decrease with larger sample sizes, consistent with expectations from the law of large numbers.

\begin{figure}[htp]
    \centering
    \setlength{\abovecaptionskip}{-0.3cm}
    \includegraphics[width=17.5cm]{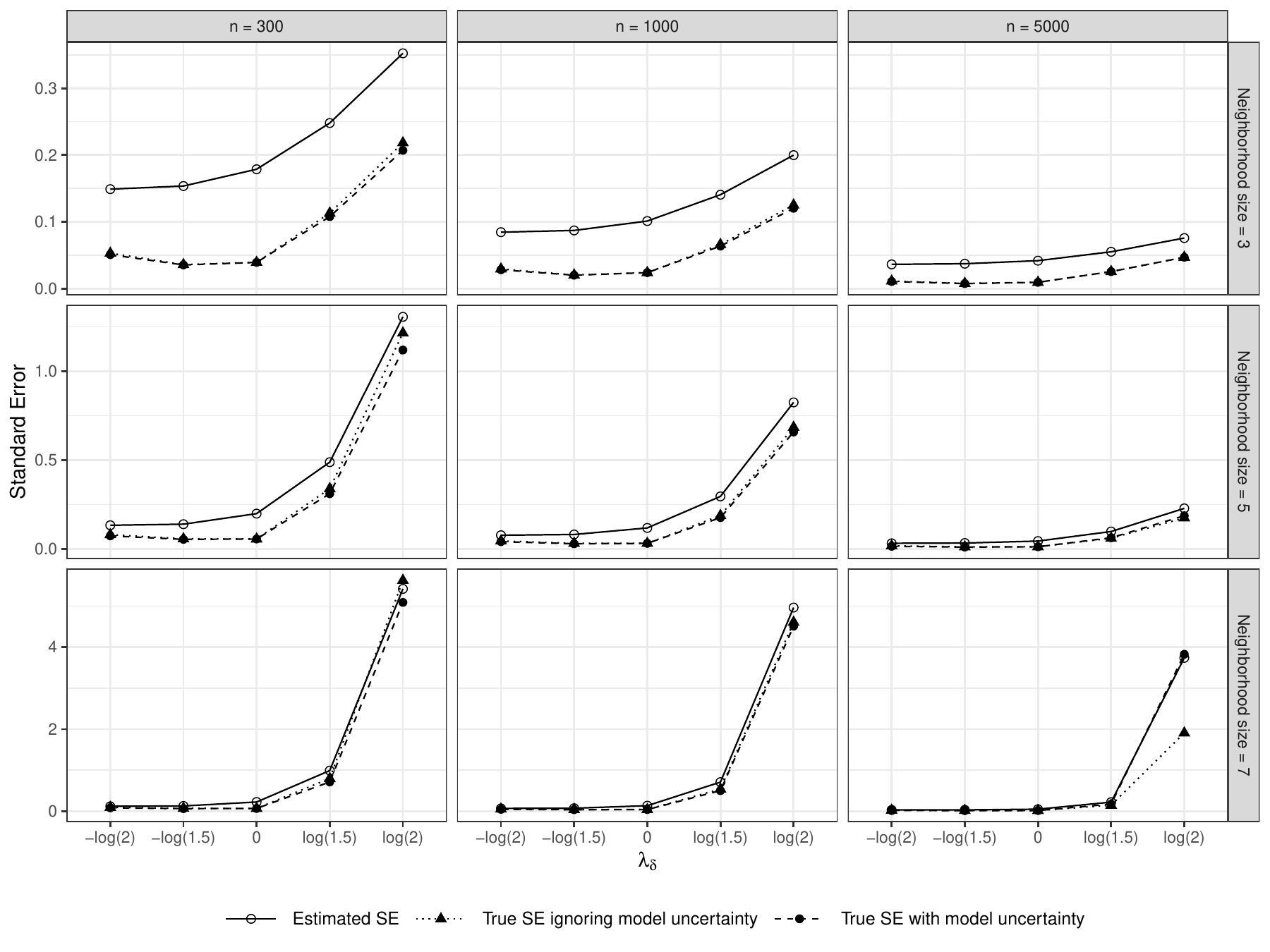}
    \caption{ Comparison of estimated and true standard errors (SEs) for $\hat{\theta}^{\delta}_{1}$.
    Each panel shows standard errors with different values of $\lambda_\delta$.  Columns correspond to sample sizes ($n = 300, 1000, 5000$), and rows correspond to exclusion neighborhood sizes ($l = 3, 5, 7$). Solid lines indicate the estimated standard errors.
    Dashed lines represent the true standard errors accounting for ERGM estimation uncertainty.
  Dotted lines represent the true standard errors assuming the fitted ERGM is the true model. }
    \label{fig:ht_sim_se}
\end{figure}

Figure~\ref{fig:ht_sim_se} compares the estimated and the two versions fo the true SEs for $\hat{\theta}_1^\delta$.  The pattern is similar to that of the H\'ajek estimator: the estimated SEs are generally larger than their true values. 

\subsection{ Simulation results under ERGMs with local dependence}
We conduct simulations under more complex ERGMs that allow for dependent edges. To ensure Assumption~\ref{asm::dependence} holds, we adopt ERGMs with local dependence as proposed by \citet{schweinberger2015local}.

\begin{definition}[Local dependence]
\label{def::local}
A random graph model exhibits local dependence if the units can be partitioned into $K \geq 2$ non-empty subsets
$\mathcal{M}_1, \ldots, \mathcal{M}_K$ such that dependencies are confined within blocks. Formally, the probability mass function of $\bm{A}$ satisfies
\begin{eqnarray}\label{model::local}
\Pr(\bm{A}=\bm{a}) = \prod_{k=1}^K  \left\{\Pr\left(A_{\mathcal{M}_k}=a_{\mathcal{M}_k}\right)
\prod_{l=1}^{k-1} \prod_{i \in \mathcal{M}_k,~j \in \mathcal{M}_l} \Pr(A_{ij}=a_{ij}) \right\},\quad a \in \mathcal{A},
\end{eqnarray}
where $A_{\mathcal{M}_k}$ denotes the adjacency matrix of the subgraph formed by all units within $\mathcal{M}_k$.
\end{definition}
Let $\bm{w}=(w_1,\ldots,w_n)$ denote block membership, where $w_i=k$ indicates that  unit $i$ belongs to  block $\mathcal{M}_k$.
We assume that both within- and between-block subnetworks follow exponential family distributions. Specifically,  the within-block subnetworks follow an ERGM of the form:
\begin{eqnarray*}
\Pr(A_{\mathcal{M}_k}=a_{\mathcal{M}_k})
\propto \exp \left\{  
\eta_{k}^T g\left( a_{\mathcal{M}_k}, {\bm X}_{\mathcal{M}_k} \right) \right\},
\end{eqnarray*}
where ${\bm X}_{\mathcal{M}_k} = \{X_i: i\in \mathcal{M}_k\}$ and $g(\cdot)$ is a vector of within-block network statistics involving  edges and covariates.
For between-block edges, we assume independent Bernoulli models with
\begin{eqnarray*}
\Pr(A_{ij}=a_{ij}) \propto \exp \left\{ \eta_B a_{ij}+ \eta_{x}^T  a_{ij}g( x_i, x_j) \right\}, \quad w_i\neq w_j.
\end{eqnarray*}
Combining the two components, the full model for $\bm{A}$ becomes
\begin{eqnarray}\label{model::ERGM_local}
\Pr( \bm{A}=\bm{a} )
\propto \exp\left\{
\sum_{k=1}^K \eta_{k}^T g \left( a_{\mathcal{M}_k}, {\bm X}_{\mathcal{M}_k} \right) + \eta_{B}  \sum_{i<j,w_i\neq w_j} a_{ij} +
\eta_{x}^T\sum_{i<j,w_i\neq w_j} a_{ij} g(x_i, x_j)
\right\}.
\end{eqnarray}

Within each block, we specify four network statistics:
\begin{eqnarray}\label{model::ERGM_local_statistics}
    g_1(\bm{a},\bm{x})&=& e(\bm{a}), \quad g_2(\bm{a},\bm{x})\ =\ \sum_{1\leq i<j\leq n}a_{ij}(x_{i1}+x_{j1}),\quad g_3(\bm{a},\bm{x})\ =\ \sum_{1\leq i<j\leq n}a_{ij}(x_{i2}+x_{j2}), \nonumber \\
    g_4(\bm{a},\bm{x}) &=& 3\sum_{i<j<k} a_{ij}a_{ik}a_{jk} + \sum_{k=2}^{n-2}(-e^{0.3})^{-(k-1)}\sum_{i<j<k}a_{ij}\binom{L_{ij}}{k},
\end{eqnarray}
where $L_{ij}=\sum_{k} a_{ik}a_{kj}$ is the number of common neighbors of units $i$ and $j$. The first three statistics match those used in the main text. The fourth statistic is the geometrically weighted edgewise shared partner statistic, which captures local clustering by aggregating triangle and $k$-triangle configurations. For the between-cluster component, we set $g(x_i, x_j) = x_i + x_j$ to capture aggregate covariate levels across blocks. We also impose structural constraints on the network by limiting the sample space $\mathcal{A}_u$ to exclude specific edges, thereby reflecting practical constraints or domain-specific knowledge.

We generate edge interventions from the ERGMs with local dependence, as specified in \eqref{model::ERGM_local} and \eqref{model::ERGM_local_statistics}. We randomly divide the sample into blocks with an average size of 50 units and adopt the same restricted sample space $\mathcal{A}_u$ as in the main text.
The data generating process follows the same procedure as in the main text, except for the definition of the exclusion neighborhoods $\mathcal{N}_i$. We modify the definition of $\mathcal{N}_i$ to include only units within the same block as unit $i$. Specifically, $\mathcal{N}_i$ consists of the $l$ nearest neighbors of unit $i$ that share the same block membership. With this specification, all units in the exclusion neighborhood belong to the same block, and the corresponding dependence neighborhood is $\mathcal{R}_i = \{j : w_j = w_i\}$, i.e.,  all units in the same block as $i$.

\begin{figure}[htp]
    \centering
    \setlength{\abovecaptionskip}{-0.1cm}
    \includegraphics[width=16cm]{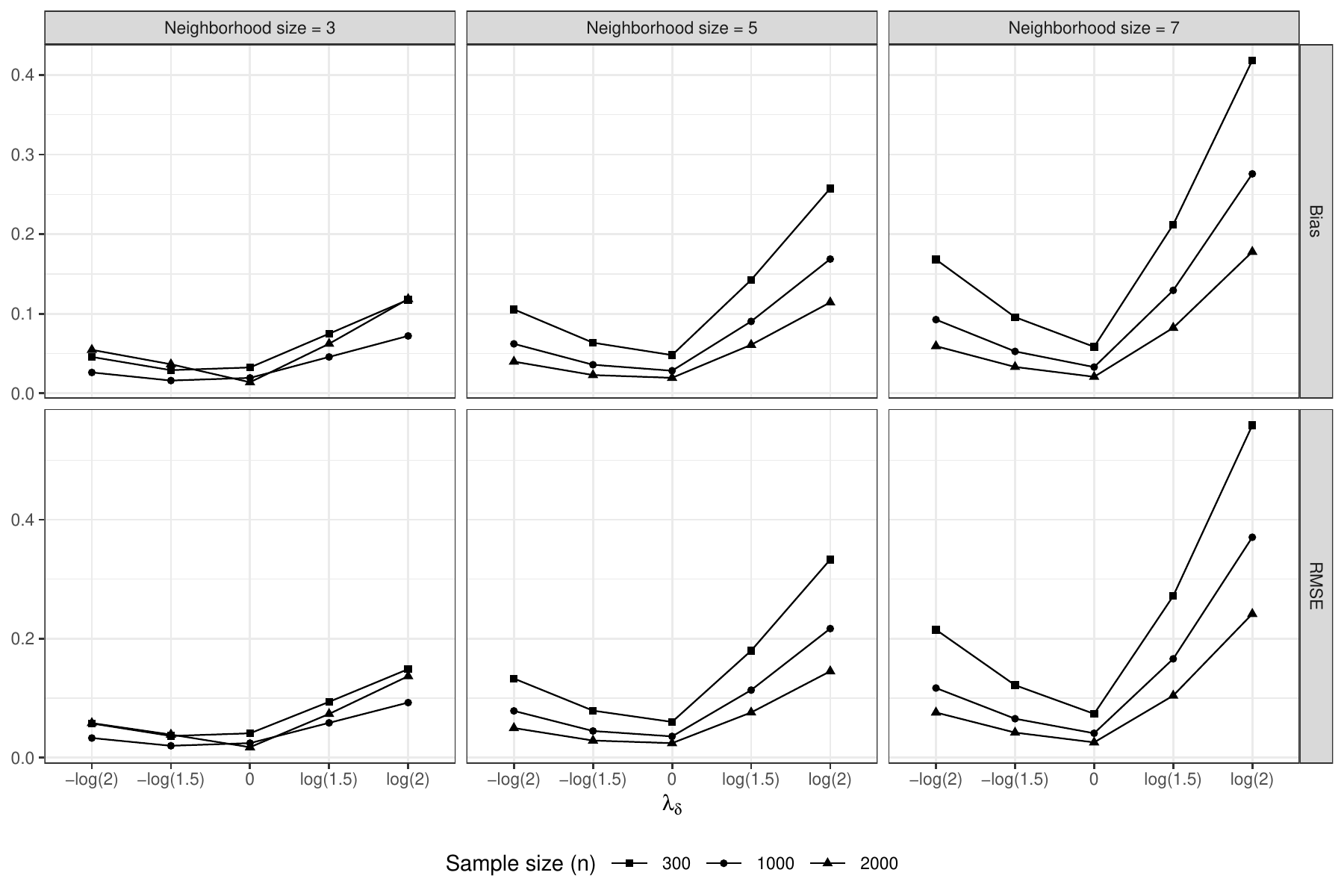}
\caption{Bias and RMSE of the Hájek estimator $\hat{\theta}^{\delta}_{2}$ under local dependence model.
The first row displays bias, and the second row displays RMSE. Each column represents a different exclusion neighborhood size $l \in \{3, 5, 7\}$.}
    \label{fig:sim_bias_rmse_local}
\end{figure}

We consider three sample sizes $n=300,1000, 2000$. 
For each scenario, we evaluate the performance of the Hájek estimator $\hat{\theta}_2^\delta$ with known models for the edges.
Figure \ref{fig:sim_bias_rmse_local} displays the bias and RMSE of the estimator. The results are consistent with those obtained under the Bernoulli random graph model in the main text.

\section{Additional details on the empirical analysis}
\label{app::application} 
\subsection{Estimation results and model diagnostics}
Table \ref{table:ergm} presents the estimated coefficients for the constrained ERGM. The results indicate that cities with higher government revenue and larger volumes of freight traffic are more likely to be connected in the railway network. In contrast, terrain relief degree is strongly negatively associated with edge formation, suggesting that unfavorable geographic conditions substantially hinder railway connectivity. For the covariate difference terms, most attributes exhibit significantly negative coefficients, indicating pronounced homophily in the network. Specifically, cities with similar levels of economic development, freight activity, population size, and terrain characteristics are significantly more likely to form direct connections.

\begin{table}[t!]
\centering
\caption{Point estimates from the constrained ERGM with standard errors in parentheses. 
``Edges'' denotes the total number of edges.  
Terms labeled ``nodecov'' represent $\sum_{1\leq i<j\leq n}a_{ij}(x_i+x_j)$ for each continuous covariate. 
Terms labeled ``absdiff'' represent $\sum_{1\leq i<j\leq n}a_{ij}|x_i-x_j|$ for continuous covariates and $\sum_{1\leq i<j\leq n}a_{ij}\bm{1}(x_i=x_j)$ for categorical covariates. }
\label{table:ergm}
\begin{threeparttable}
\begin{tabular}{lllll}
\toprule
\textbf{Model Term} & \textbf{Coefficient} & \vline\vline &\textbf{Model Term} & \textbf{Coefficient} \\
\cline{1-2}\cline{4-5}
Edges & ~-0.590 (0.160)*** & \vline\vline &  absdiff.Income & ~-0.130 (0.126) \\
nodecov.GDP\_per\_capita & ~-0.049 (0.052) & \vline\vline &absdiff.Freight\_traffic & ~-0.431 (0.099)*** \\
nodecov.Income & ~~0.304 (0.125)* &\vline\vline & absdiff.Population & ~-0.299 (0.070)*** \\
nodecov.Freight\_traffic &~~0.397(0.090)*** &\vline\vline & absdiff.Terrain & ~-1.670 (0.166)*** \\
nodecov.Population & ~-0.128 (0.051)* & \vline\vline &nodematch.Airport  &~ 0.065 (0.085) \\
nodecov.Terrain & ~-0.199 (0.062)**& \vline\vline &nodematch.Capital & ~-0.897 (0.113)*** \\
absdiff.GDP\_per\_capita & ~-0.361 (0.067)***  &\vline\vline &   \\
\bottomrule
\end{tabular}
\begin{tablenotes}
\small
\item Note: *** $p<0.001$, ** $p<0.01$, * $p<0.05$. Standard errors in parentheses.
\end{tablenotes}
\end{threeparttable}
\end{table}

We assess the goodness of fit of the constrained ERGM by following the procedure outlined in \citet{hunter2008goodness}. Specifically, we generate 100 networks from the fitted model and compare the distributions of the network statistics and node degree with the observed values. 

\begin{figure}[!h]
    \centering
    \setlength{\abovecaptionskip}{0.1cm}
    \includegraphics[width=1\linewidth]{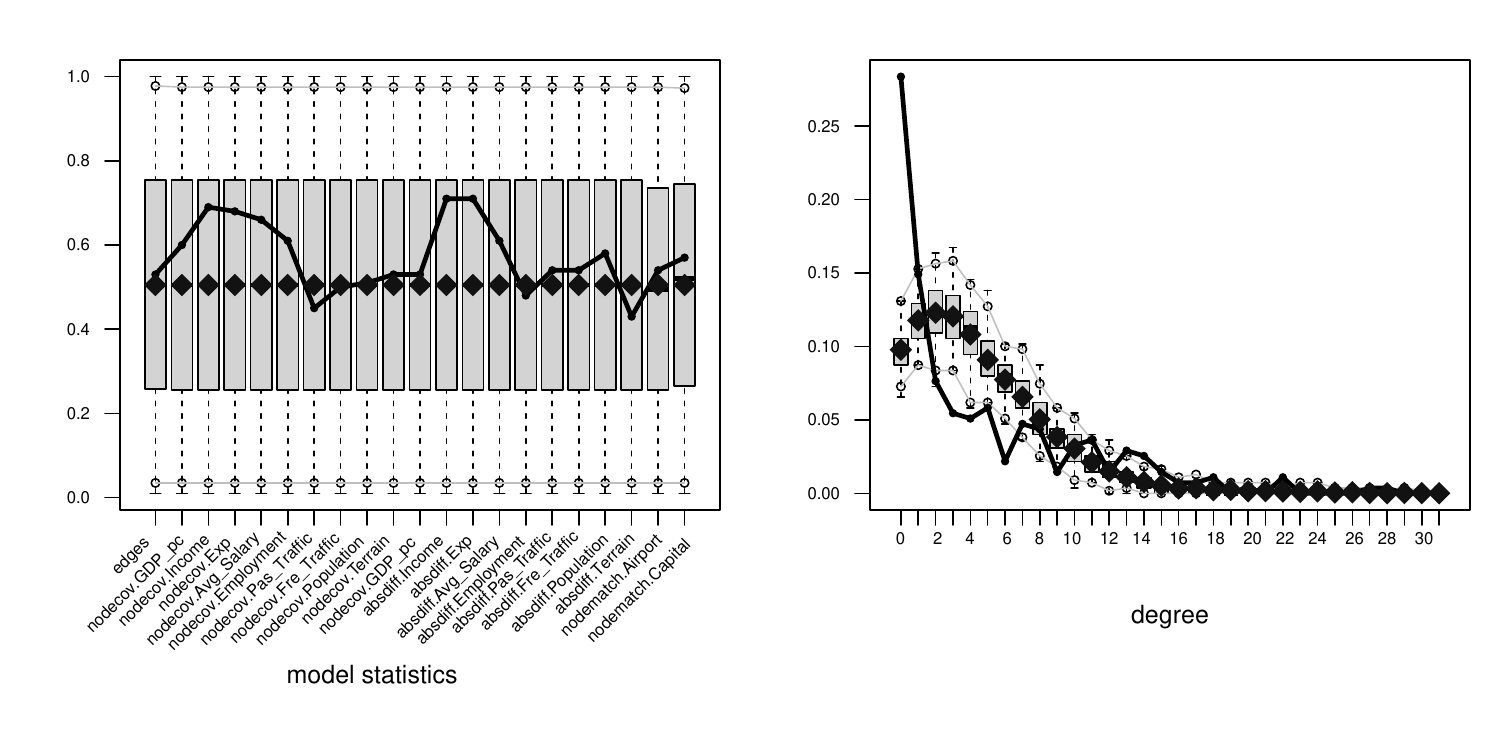}
    \caption{Goodness-of-fit diagnostics for the constrained ERGM with local dependence. The left panel compares the observed network statistics with those from 100 simulated networks. The grey boxplots show the distribution of each statistic across the simulated networks (standardized to the interval $[0,1]$), and the black line indicates the percentile of the observed statistic within this distribution. The right panel compares the observed degree distribution with that from the 100 simulated networks, using the same boxplot–percentile display as in the left panel.}
    \label{fig:model_diagnostics}
\end{figure}

Figure \ref{fig:model_diagnostics} summarizes the results. In the left panel, the observed network statistics lie close to the medians of those from the 100 generated networks. In the right panel, the observed degree distribution does not deviate substantially from the simulated distributions. Taken together, these results indicate that the model provides a reasonable fit to the observed network

\subsection{Sensitivity analysis}
\begin{figure}[!h]
   \centering
    \setlength{\abovecaptionskip}{-0.3cm}
    \includegraphics[width=0.9\linewidth]{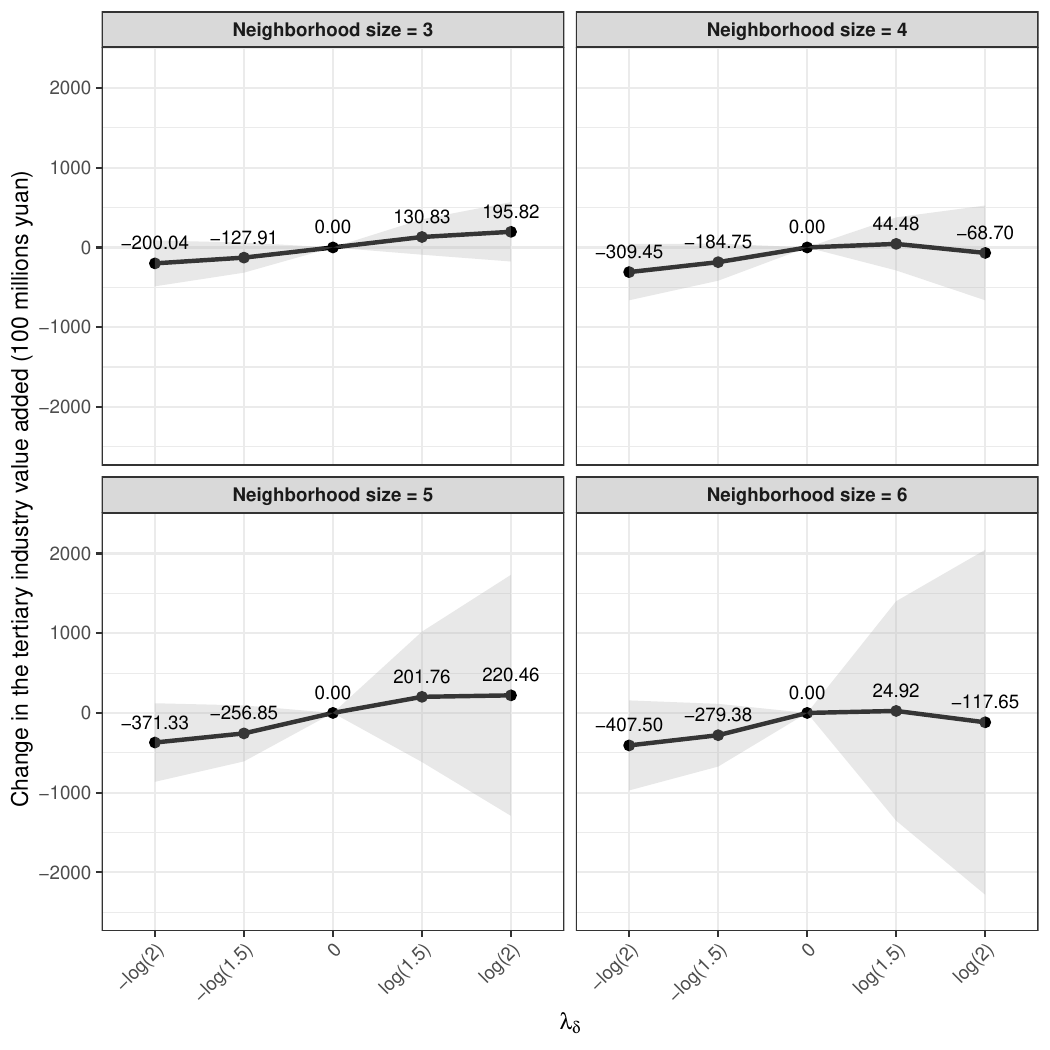}
    \caption{ The impact of China's railway network on the tertiary industry value added under a 1.5-hour travel-time cutoff. Each panel displays results on for a fixed exclusion neighborhood size ($l \in {3, 4, 5, 6}$). The lines show the estimates of $\theta^\delta-\theta^0$ across varying levels of intervention intensity ($\lambda_\delta$), and the shaded regions indicate 95\% pointwise confidence intervals.  }
    \label{fig:result_gdp_1.5hour}
\end{figure}

\begin{figure}[htp]
   \centering
    \setlength{\abovecaptionskip}{-0.3cm}
    \includegraphics[width=0.9\linewidth]{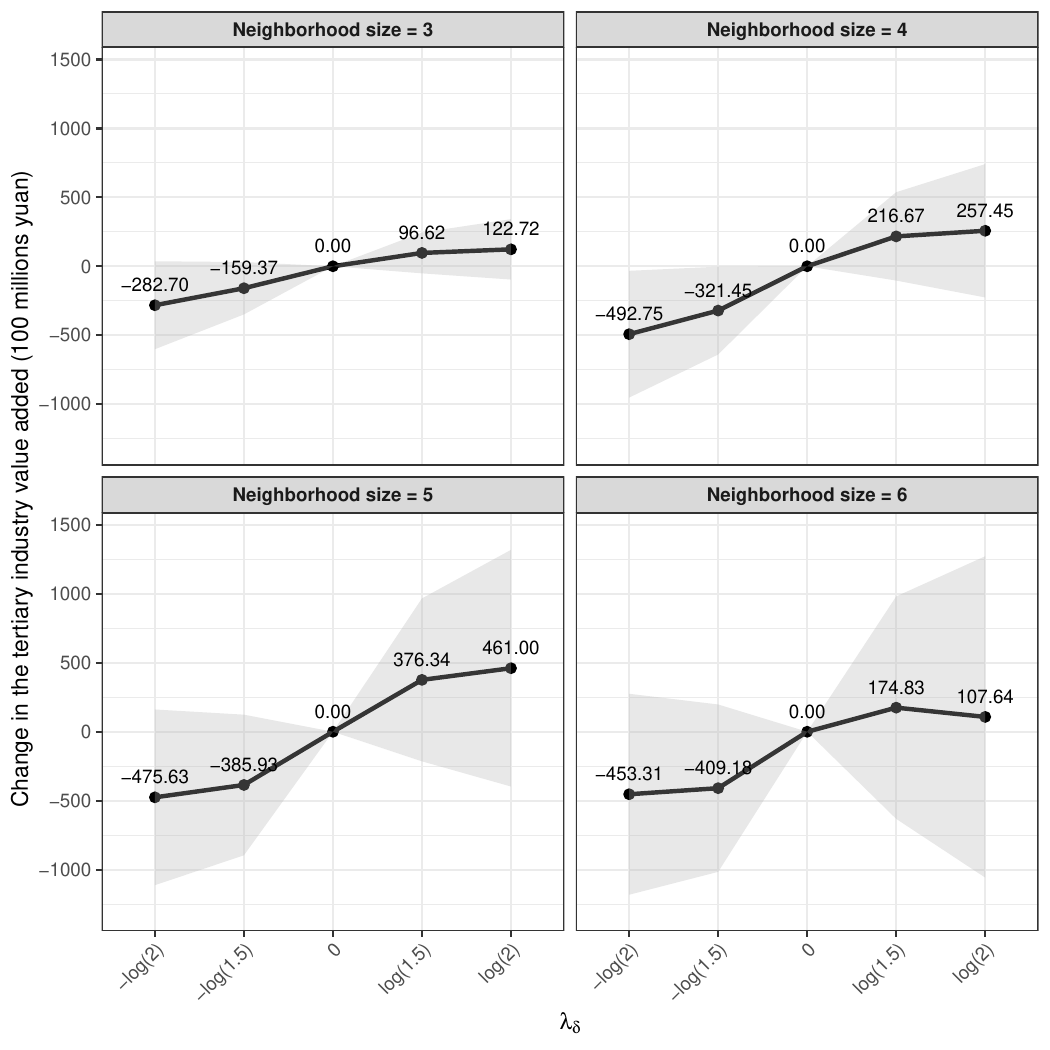}
    \caption{ The impact of China's railway network on the tertiary industry value added under a 2.5-hour travel-time cutoff. Each panel displays results on for a fixed exclusion neighborhood size ($l \in {3, 4, 5, 6}$). The lines show the estimates of $\theta^\delta-\theta^0$ across varying levels of intervention intensity ($\lambda_\delta$), and the shaded regions indicate 95\% pointwise confidence intervals.   }
    \label{fig:result_gdp_2.5hour}
\end{figure}

We conduct sensitivity analysis to evaluate the robustness of our findings by varying the direct-travel time cutoff. In addition to the 2-hour threshold used in the main analysis, we consider alternative cutoffs of 1.5 hours and 2.5 hours, corresponding to distance thresholds of $d_u = 600$ km and $d_u = 1000$ km, respectively. Figures~\ref{fig:result_gdp_1.5hour} and~\ref{fig:result_gdp_2.5hour} present the estimation results under these alternative specifications. Under both scenarios, the qualitative patterns are broadly consistent with those observed in the main analysis.


\end{document}